\documentclass{article} 
\usepackage[pdftex]{graphicx} 
\usepackage[utf8]{inputenc} 
\usepackage{amsmath} 
\usepackage{amssymb} 
\usepackage{amstext} 
\usepackage[mathscr]{eucal} 
\usepackage{array} 
\numberwithin{equation}{section} 

\DeclareMathOperator{\arth}{artanh}

\newcommand{\sfig}[5]{\begin{figure}[!htb]\centering\includegraphics[angle=#1,scale=#2]{#3}\label{fig:#4}\caption{#5}\end{figure}} 
 
\newcommand{\eq}[2]{\begin{eqnarray}#1\label{eq:#2}\end{eqnarray}} 
\newcommand{\eqd}[4]{\begin{eqnarray}#1\label{eq:#2}\\#3\label{eq:#4}\end{eqnarray}} 
\newcommand{\eqt}[6]{\begin{eqnarray}#1\label{eq:#2}\\#3\label{eq:#4}\\#5\label{eq:#6}\end{eqnarray}} 
\newcommand{\eqq}[8]{\begin{eqnarray}#1\label{eq:#2}\\#3\label{eq:#4}\\#5\label{eq:#6}\\#7\label{eq:#8}\end{eqnarray}} 

\begin{document} 
 
\title{Accelerating expansion in the swisscheese model} 
\author{Gyula Bene and Adelinda Csapó\\ 
Institute of Physics, Eötvös University\\ 
Pázmány P. s. 1/A, H-1117 Budapest, Hungary 
} 
\maketitle 
\begin{abstract} 
A version of the Swiss-cheese model is investigated. The flat Friedmann-Robertson-Walker (FRW) universe is modified by the addition of several spherical regions with Lema\^itre-Tolman-Bondi metric.  
   
We discuss light propagation in this model in detail to pave the way for a detailed numerical study of the Hubble diagram.
\end{abstract}

\section{Introduction} 
 
Type Ia supernova data interpreted in terms of homogeneous, isotropic cosmological models necessitate the introduction of a huge amount of dark energy, i.e., a substance of negative pressure (like the cosmological constant). A recent alternative suggestion is the consideration of inhomogenity in the density distribution. Indeed, lensing effects of local matter abundances modify the positions of distant objects on the Hubble diagram. There are several recent papers about this question without a concordance whether this effect is strong enough to give account of supernova data,[] albeit the majority of the authors seems to deny this possibility. The question is rather important since the existence of dark energy (or cosmological constant) profoundly modifies our knowledge about matter, already challenged by the existence of dark matter.       
 
In the present paper we consider the problem within the framework of the exactly solvable Swiss-cheese model. The version of the model we use is constructed the following way. 
 
Nonoverlapping spheres are cut from 
a flat Friedmann-Robertson-Walker (FRW) universe. The mass they contained before is compressed within each 
sphere to a smaller sphere with homogeneous density distribution. Hence the 
inner spheres form sections of some closed FRW model. Between the outer and 
inner spheres there is a vacuum, where, due to spherical symmetry,  
the Schwarzschield metric describes the gravitational field. Within the inner 
spheres the closed FRW metric is valid, while outside the cut spheres the flat 
FRW metric is relevant. The metric and its first derivatives are 
continuous across the bordering surfaces of the different regions.

We use the Landau conventions, i.e., we assume $+---$ signature for the metric.  
The zeroth component of a four-vector is timelike,  
the first, second and third components are spacelike. Four-vectors are indexed  
by Latin letters, 
three-vectors by Greek letters. Throughout we use $c=1$ units. At light propagation we use the index $1$ for labeling initial quantities and the index $0$ for labeling final (i.e., present) quantities. 
\section{Light propagation in flat FRW universe} 
 
Light propagation in gravitational field is governed by the Sachs optical equations reviewed below.  
 
A light ray may be parametrized along its path by a parameter $\lambda$ so 
that its four-velocity $u^i$ is defined by 
\begin{eqnarray} 
u^i=\frac{dx^i}{d\lambda}\;.\label{he1.1} 
\end{eqnarray} 
It is a null vector, 
\begin{eqnarray} 
u^iu_i=0\label{he1.2} 
\end{eqnarray} 
and satisfies the geodesic equation 
\begin{eqnarray} 
u^i_{;k}u^k=0\;.\label{he1.3} 
\end{eqnarray} 
Let us introduce another independent null vector $w^i$ along the path.  
It satisfies 
\begin{eqnarray} 
w^iw_i=0\;,\label{he1.4} 
\end{eqnarray} 
\begin{eqnarray} 
w^iu_i=1\label{he1.5} 
\end{eqnarray} 
and 
\begin{eqnarray} 
w^i_{;k}u^k=0\;.\label{he1.6} 
\end{eqnarray} 
Furthermore, let us define two spacelike unit vectors $L^i_\alpha$ ($\alpha, 
\beta=1,2$)  
that are orthogonal to both the null 
vectors $u^i$ and $w^i$: 
\begin{eqnarray} 
L^i_\alpha u_i=0\label{he1.7} 
\end{eqnarray} 
\begin{eqnarray} 
L^i_\alpha w_i=0\label{he1.8} 
\end{eqnarray} 
\begin{eqnarray} 
L^i_\alpha L_{i\beta}=-\delta_{\alpha \beta}\label{he1.9} 
\end{eqnarray} 
We also require that the vectors $L^i_\alpha$ are parallel translated along 
the light ray:  
\begin{eqnarray} 
L^i_{\alpha;k}u^k=0\label{he1.10} 
\end{eqnarray} 
Eqs. (\ref{he1.3}), (\ref{he1.6}), (\ref{he1.10}) ensure that if 
Eqs. (\ref{he1.4}), (\ref{he1.5}), (\ref{he1.7}), (\ref{he1.8}), (\ref{he1.9}) are 
satisfied at one single point of the light ray, they will be satisfied at any 
other point of it as well. 
For the separation $\xi^i$ of two light rays we assume 
\begin{eqnarray} 
\xi^iu_i=0\label{he1.11} 
\end{eqnarray} 
and 
\begin{eqnarray} 
\xi^iw_i=0\;,\label{he1.12} 
\end{eqnarray} 
hence we have 
\begin{eqnarray} 
\xi^i=\sum_{\alpha=1,2}d_\alpha L^i_\alpha\;.\label{he1.13} 
\end{eqnarray} 
The coefficients $d_\alpha$ describe the proper separation of the two 
nearby light rays. For their derivative with respect to the path parameter 
$\lambda$ we have in the most general case (since no rotation is possible) 
\begin{eqnarray} 
\frac{d d_\alpha}{d\lambda} = \sum_{\beta=1,2} \left(\Theta \delta_{\alpha \beta}+w_{\alpha \beta}\right)d_\beta\label{he1.14} 
\end{eqnarray}   
where $w_{\alpha \beta}$ is symmetric and traceless: 
\begin{eqnarray} 
w_{\alpha \beta}=\left(\begin{array}{cc}\rho&\sigma \\ \sigma&-\rho\end{array}\right)\;.\label{he1.15} 
\end{eqnarray} 
Now, the geodesic deviation equation 
\begin{eqnarray} 
\frac{D^2\xi^i}{D\lambda^2}=R^i_{jkl}u^iu^k\xi^l\label{he1.16} 
\end{eqnarray} 
leads to the Sachs optical equations for the expansion rate $\Theta$ and shear 
$w_{\alpha \beta}$: 
\begin{eqnarray} 
\frac{d\Theta}{d\lambda}+\Theta^2+\frac{1}{2}w^2=-\frac{1}{2}R_{jk}u^ju^k\label{he1.17} 
\end{eqnarray} 
\begin{eqnarray} 
\frac{dw_{\alpha\beta}}{d\lambda}+2\Theta w_{\alpha\beta} =C_{ ijkl}L^i_\alpha 
u^ju^kL^l_\beta\label{he1.18} 
\end{eqnarray} 
where 
\begin{eqnarray} 
w^2=w_{\alpha\beta}w_{\beta\alpha}\label{he1.19} 
\end{eqnarray}  
and $C_{ ijkl}$ stands for the Weyl tensor. 
 
The physical situation (e.g., that the two nearby light rays are 
emitted from a point source) is specified by the initial conditions. After having solved these 
equations, $\Theta$ governs the change of the cross section $A$ of a light 
beam: 
\begin{eqnarray} 
\frac{dA}{d\lambda}=2\Theta A\;.\label{he1.20} 
\end{eqnarray} 
 
In a flat FRW universe the metric can be written as 
\eq{ds^2=dt^2-a(t)^2\left(dr^2+r^2\left(d\vartheta^2+\sin^2\vartheta d\varphi^2\right)\right)\quad.}{he1.0}
In a matter dominated universe the scale factor $a(t)$ may be given by
\eq{a(t)&=&a(t_0)\left(\frac{3}{2}H_0t\right)^{\frac{2}{3}}}{ef.00a}
where 
\eq{H_0^2=\frac{8\pi G}{3}\rho(t_0)}{ef.00xb} 
according to the first Friedmann equation, $H_0$ denoting the present value of Hubble's parameter. 

The following formulae hold, provided that the light is propagating radially away from the origin. 
 
The four-velocity is given by (up to a constant multiplier, which determines the color of the light at emission) 
\begin{eqnarray} 
u^{t}&=&\frac{1}{a}\label{e3.19}\\ 
u^{r}&=&\frac{1}{a^2}\label{e3.20}\\ 
u^{\vartheta}&=&0\\ 
u^{\varphi}&=&0\label{e3.21} 
\end{eqnarray} 
For the trajectory we get 
\begin{eqnarray} 
\dot r=\frac{1}{a}\label{e3.22} 
\end{eqnarray} 
or 
\begin{eqnarray} 
r(t)=\int_{t_1}^{t} \frac{dt}{a}=3\left(\frac{t}{a(t)}-\frac{t_1}{a(t_1)}\right)\label{e3.23} 
\end{eqnarray}

For the expansion rate we have 
\begin{eqnarray} 
\Theta=\frac{\dot a}{a^2}+\frac{1}{a^2 r}=\frac{\dot{(a r)}}{a^2 r}\label{e3.24} 
\end{eqnarray} 
and thus the cross section area of the beam is 
\begin{eqnarray} 
A=\Omega a^2r^2\label{e3.25} 
\end{eqnarray} 
The redshift is defined by 
\eq{z=\frac{u^{t}(t_1)}{u^{t}(t_0)}-1 
=\frac{a(t_0)}{a(t_1)}-1\quad,}{2.1.87} 
For the intensity we have 
\eq{I=L\frac{\Omega}{4\pi}\frac{1}{A(t_0)(1+z)^2}\quad,}{2.1.86a} 
where 
\eq{A(t_0)=\Omega\left(\frac{2}{H_0}\left(1-\frac{1}{\sqrt{1+z}}\right)\right)^2\quad}{2.1.86b} 
and $L$ stands for the absolute luminosity. 
Then the luminosity distance is given by 
\eq{d_L=\sqrt{\frac{L}{4\pi I}}=\frac{2}{H_0}\sqrt{1+z}\left(\sqrt{1+z}-1\right)\quad.}{2.1.90.1} 
This is the Hubble-diagram in a matter-dominated flat FRW universe. 
\sfig{-90}{0.45}{flatFRW.jpg}{hubb-flat}{The Hubble-diagram in a matter-dominated flat FRW universe.}\\ 
 
 
Let us describe light propagation in an arbitrary direction. As space is homogeneous and isotropic, it is enough to perform a translation of the origin. This means that  
\begin{eqnarray} 
{\bf r}={\bf r}'+{\bf r}_k 
\end{eqnarray} 
 
Suppose that in the coordinate system K' the light ray is radial.  
 
The above coordinate transformation implies that 
\begin{eqnarray} 
r'&=&\sqrt{r^2+r_k^2-2rr_k\left(\cos\vartheta\cos\vartheta_k+\sin\vartheta\sin\vartheta_k\cos(\varphi-\varphi_k)\right)}\qquad\label{tr0a}\\ 
\cos\vartheta'&=&\frac{r\cos\vartheta-r_k\cos\vartheta_k}{\sqrt{r^2+r_k^2-2rr_k\left(\cos\vartheta\cos\vartheta_k+\sin\vartheta\sin\vartheta_k\cos(\varphi-\varphi_k)\right)}}\\ 
\tan \varphi'&=&\frac{r\sin\vartheta\sin\varphi-r_k\sin\vartheta_k\sin\varphi_k}{r\sin\vartheta\cos\varphi-r_k\sin\vartheta_k\cos\varphi_k} \label{tr0}
\end{eqnarray}  

The trajectory is given by the inverse coordinate transformation 
 
\begin{eqnarray} 
r&=&\sqrt{r'^2+r_k^2+2r'r_k\left(\cos\vartheta'\cos\vartheta_k+\sin\vartheta'\sin\vartheta_k\cos(\varphi'-\varphi_k)\right)}\label{tr1a}\\ 
\cos\vartheta&=&\frac{r'\cos\vartheta'+r_k\cos\vartheta_k}{\sqrt{r'^2+r_k^2+2r'r_k\left(\cos\vartheta'\cos\vartheta_k+\sin\vartheta'\sin\vartheta_k\cos(\varphi'-\varphi_k)\right)}}\qquad\label{tr1b}\\ 
\tan \varphi&=&\frac{r'\sin\vartheta'\sin\varphi'+r_k\sin\vartheta_k\sin\varphi_k}{r'\sin\vartheta'\cos\varphi'+r_k\sin\vartheta_k\cos\varphi_k} \label{tr1}
\end{eqnarray}  
where $\vartheta'$ and $\varphi'$ are constant, while 
\begin{eqnarray} 
r'=3\left(\frac{t}{a(t)}-\frac{t_1}{a(t_1)}\right)\;. \label{tr2}
\end{eqnarray} 

The expansion rate is
\eq{\Theta=\frac{\dot a}{a^2}+\frac{1}{a^2 r'}=\frac{\dot{(a r')}}{a^2 r'}}{sw_f.3}
and for the cross section we have
\eq{A=\Omega a^2r'^2}{sw_f.4} 
 
Let us determine the transformation of the velocity. The timelike component is clraly unchanged. Covariant spatial vectorcomponents transform as  
\begin{eqnarray} 
v_\alpha=\frac{\partial x'^\beta}{\partial x^\alpha}v'_\beta \label{eq:xxx1}
\end{eqnarray} 
The transformation matrix elements are given explicitly by 
\begin{eqnarray} 
\frac{\partial r'}{\partial r}&=&\frac{r-r_k\left(\cos\vartheta\cos\vartheta_k+\sin\vartheta\sin\vartheta_k\cos(\varphi-\varphi_k)\right)}{\sqrt{r^2+r_k^2-2rr_k\left(\cos\vartheta\cos\vartheta_k+\sin\vartheta\sin\vartheta_k\cos(\varphi-\varphi_k)\right)}}\nonumber\\ 
\frac{\partial r'}{\partial \vartheta}&=&\frac{rr_k\left(\sin\vartheta\cos\vartheta_k-\cos\vartheta\sin\vartheta_k\cos(\varphi-\varphi_k)\right)}{\sqrt{r^2+r_k^2-2rr_k\left(\cos\vartheta\cos\vartheta_k+\sin\vartheta\sin\vartheta_k\cos(\varphi-\varphi_k)\right)}}\nonumber\\ 
\frac{\partial r'}{\partial \varphi}&=&\frac{rr_k\sin\vartheta\sin\vartheta_k\sin(\varphi-\varphi_k)}{\sqrt{r^2+r_k^2-2rr_k\left(\cos\vartheta\cos\vartheta_k+\sin\vartheta\sin\vartheta_k\cos(\varphi-\varphi_k)\right)}}\nonumber\\ 
\frac{\partial \vartheta'}{\partial r}&=&\frac{-\left(r\cos\vartheta_k+r_k\cos\vartheta\right)+\left(r\cos\vartheta+r_k\cos\vartheta_k\right)\left(\cos\vartheta\cos\vartheta_k+\sin\vartheta\sin\vartheta_k\cos(\varphi-\varphi_k)\right)}{\left(r^2+r_k^2-2rr_k\left(\cos\vartheta\cos\vartheta_k+\sin\vartheta\sin\vartheta_k\cos(\varphi-\varphi_k)\right)\right)}\nonumber\\ 
&&\times\frac{r_k}{\sqrt{r^2\sin^2\vartheta+r_k^2\sin^2\vartheta_k-2rr_k\sin\vartheta\sin\vartheta_k\cos(\varphi-\varphi_k)}}\nonumber\\ 
\frac{\partial \vartheta'}{\partial \vartheta}&=&\left[\sin\vartheta+\frac{r_k\left(r\cos\vartheta-r_k\cos\vartheta_k\right)\left(\sin\vartheta\cos\vartheta_k-\cos\vartheta\sin\vartheta_k\cos(\varphi-\varphi_k)\right)}{\left(r^2+r_k^2-2rr_k\left(\cos\vartheta\cos\vartheta_k+\sin\vartheta\sin\vartheta_k\cos(\varphi-\varphi_k)\right)\right)}\right]\nonumber\\ 
&&\times\frac{r}{\sqrt{r^2\sin^2\vartheta+r_k^2\sin^2\vartheta_k-2rr_k\sin\vartheta\sin\vartheta_k\cos(\varphi-\varphi_k)}}\nonumber\\ 
\frac{\partial \vartheta'}{\partial \varphi}&=&\frac{r\left(r\cos\vartheta-r_k\cos\vartheta_k\right)\sin\vartheta\sin\vartheta_k\sin(\varphi-\varphi_k)}{\left(r^2+r_k^2-2rr_k\left(\cos\vartheta\cos\vartheta_k+\sin\vartheta\sin\vartheta_k\cos(\varphi-\varphi_k)\right)\right)}\nonumber\\ 
&&\times\frac{r_k}{\sqrt{r^2\sin^2\vartheta+r_k^2\sin^2\vartheta_k-2rr_k\sin\vartheta\sin\vartheta_k\cos(\varphi-\varphi_k)}}\nonumber\\ 
\frac{\partial \varphi'}{\partial r}&=&-\frac{r_k\sin\vartheta\sin\vartheta_k\sin(\varphi-\varphi_k)}{r^2\sin^2\vartheta+r_k^2\sin^2\vartheta_k-2rr_k\sin\vartheta\sin\vartheta_k\cos(\varphi-\varphi_k)}\nonumber\\ 
\frac{\partial \varphi'}{\partial \vartheta}&=&-\frac{rr_k\cos\vartheta\sin\vartheta_k\sin(\varphi-\varphi_k)}{r^2\sin^2\vartheta+r_k^2\sin^2\vartheta_k-2rr_k\sin\vartheta\sin\vartheta_k\cos(\varphi-\varphi_k)}\nonumber\\ 
\frac{\partial \varphi'}{\partial \varphi}&=&\frac{r^2\sin^2\vartheta-rr_k\sin\vartheta\sin\vartheta_k\cos(\varphi-\varphi_k)}{r^2\sin^2\vartheta+r_k^2\sin^2\vartheta_k-2rr_k\sin\vartheta\sin\vartheta_k\cos(\varphi-\varphi_k)}\nonumber 
\end{eqnarray} 
 
The transformation of contravariant vectors may be derived by pulling indices down, transforming the covariant vector thus obtained, and finally pulling indices up. Note that the metric is the same both before and after the transformation, if expressed in terms of the corresponding coordinates. Hence for the four-velocity we get 
\begin{eqnarray} 
u^{t}&=&\frac{1}{a}\label{flat_ut}\\ 
u^{r}&=&\frac{r-r_k\left(\cos\vartheta\cos\vartheta_k+\sin\vartheta\sin\vartheta_k\cos(\varphi-\varphi_k)\right)}{a^2\sqrt{r^2+r_k^2-2rr_k\left(\cos\vartheta\cos\vartheta_k+\sin\vartheta\sin\vartheta_k\cos(\varphi-\varphi_k)\right)}}\label{flat_ur}\\ 
u^{\vartheta}&=&\frac{r_k\left(\sin\vartheta\cos\vartheta_k-\cos\vartheta\sin\vartheta_k\cos(\varphi-\varphi_k)\right)}{a^2r\sqrt{r^2+r_k^2-2rr_k\left(\cos\vartheta\cos\vartheta_k+\sin\vartheta\sin\vartheta_k\cos(\varphi-\varphi_k)\right)}}\label{flat_utheta}\\ 
u^{\varphi}&=&\frac{r_k\sin\vartheta_k\sin(\varphi-\varphi_k)}{a^2r\sin\vartheta\sqrt{r^2+r_k^2-2rr_k\left(\cos\vartheta\cos\vartheta_k+\sin\vartheta\sin\vartheta_k\cos(\varphi-\varphi_k)\right)}}\qquad\qquad\label{flat_uphi} 
\end{eqnarray}

\section{Light propagation in closed FRW universe} 
In a closed FRW universe the metric can be written as 
\eq{ds^2=dt^2-a(t)^2\left(\frac{dr^2}{1-r^2}+r^2\left(d\vartheta^2+\sin^2\vartheta d\varphi^2\right)\right)\quad.}{fc.00} 
In a matter dominated universe the scale factor $a(t)$ may be given parametrically: 
\eqd{a(t)&=&\beta\sin^2\chi}{fc.00a}{t&=&\beta\left(\chi-\sin\chi\cos\chi\right)}{fc.0xa} 
where 
\eq{\beta=\frac{8\pi G}{3}\rho(t_0) a^3(t_0)\quad.}{fc.0xb} 
Note that
according to the first Friedmann equation
\eq{\frac{\beta}{a^3(t_0)} =H_0^2+\frac{1}{a^2(t_0)}\quad,}{fc.0xc} 
 
 $H_0$ denoting the present value of Hubble's parameter (which may deviate from that in the flat FRW region). Further, we have

\eq{ \dot \chi=\frac{1}{2\beta\sin^2\chi}\quad.}{fc.0xd}

Provided that the light propagates radially in the chosen coordinate system, 
 $d\vartheta=0$ and $d\varphi=0$ holds, hence $ds^2=0$ implies 
\eq{dt^2=a(t)^2\frac{dr^2}{1-r^2}\quad.}{fc.01a} 
The solution is 
\eq{r=\sin\left(2(\chi-\chi_1)\right)\quad,}{fc.01b} 
where $\chi_1=\chi(t_1)$.  
 
The components of the four velocity are (up to a constant multiplier again)
\eqq{u^{t}&=&\frac{1}{a}}{fc.15} 
{u^{r}&=&\frac{\sqrt{1-r^2}}{a^2}}{fc.16} 
{u^{\vartheta}&=&0}{fc.17} 
{u^{\varphi}&=&0}{fc.18} 
 
The cross section area of the light beam emanated from the origin may be expressed as  
\eq{A=\Omega a^2\;r^2\quad,}{fc.k.1} 
which implies according to Eq.(\ref{he1.20}) that 
\eq{\Theta=\frac{\dot{a}}{a^2}+\frac{\sqrt{1-r^2}}{a^2\;r}\quad.}{fc.k.2} 
Then 
\eq{\dot{\Theta}+a\Theta^2=-\frac{3}{2}\frac{\beta}{a^4}\quad,}{fc.k.3} 
i.e., it satisfies Sach's optical equations (\ref{he1.17}), (\ref{he1.18}).  
 
The redshift is given by 
\eq{z=\frac{a(t_0)}{a(t_1)}-1=\frac{\sin^2(\chi_0)}{\sin^2(\chi_1)}-1\quad,}{fc.k.4} 
For the luminosity distance we have 
\eq{d_L=(1+z)\sqrt{\frac{A}{\Omega}}=(1+z)a(t_0)r_0=\beta\frac{\sin^4(\chi_0)}{\sin^2(\chi_1)}\sin\left(2(\chi_0-\chi_1)\right)\quad.}{fc.k.6} 
Eqs.(\ref{eq:fc.k.4}), (\ref{eq:fc.k.6}) express the Hubble diagram in parametric form, the parameter being $\chi_1$. Explicitly, using Eqs.(\ref{eq:fc.00a}), (\ref{eq:fc.0xc}), (\ref{eq:fc.01b}), too, we get for the expanding case ($\chi_0<\pi/2$)
\eq{d_L&=&\frac{2}{H_0}\left[\frac{H_0^2a_0^2\left((1+z)\left(1+H_0^2a_0^2\right)-2\right)}{\left(1+H_0^2a_0^2\right)^2}\right.\nonumber\\
&&\left.+\frac{H_0a_0\left(1-H_0^2a_0^2\right)\sqrt{z+(1+z)H_0^2a_0^2}}{\left(1+H_0^2a_0^2\right)^2}\right]  }{fc.k.6v}
\newpage
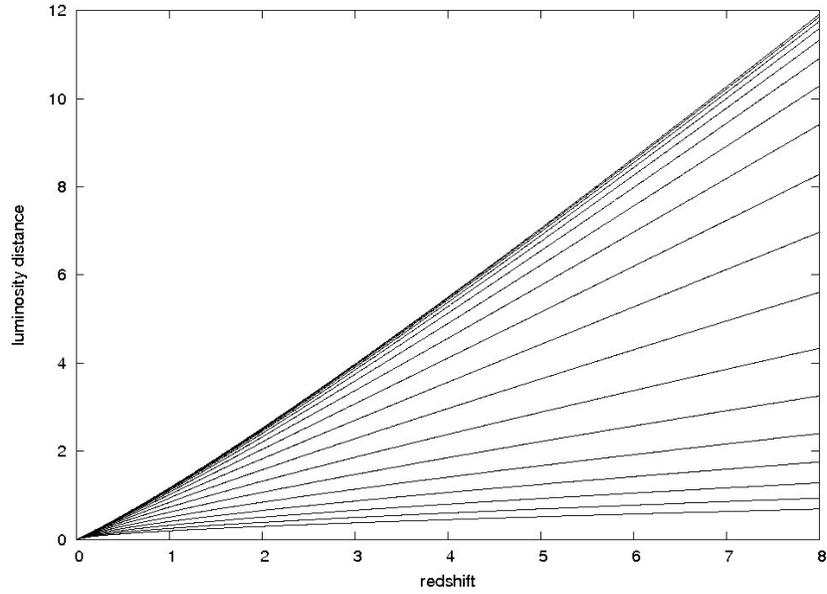
\sfig{-90}{0.45}{closedexpandingFRW.jpg}{hubb-closedexpanding}{The Hubble diagram in a closed, expanding FRW universe. Different curves correspond to different $H_0a_0$ values. Note that in the limit $H_0a_0\gg 1$ the Hubble diagram of the flat FRW is obtained.}
In the collapsing case ($\pi>\chi_0>\pi/2$) the parametric form is more suitable, since the Hubble diagram becomes multivalued, because light may reach the observer from the opposite direction, through the rest of the universe as well.
\newpage
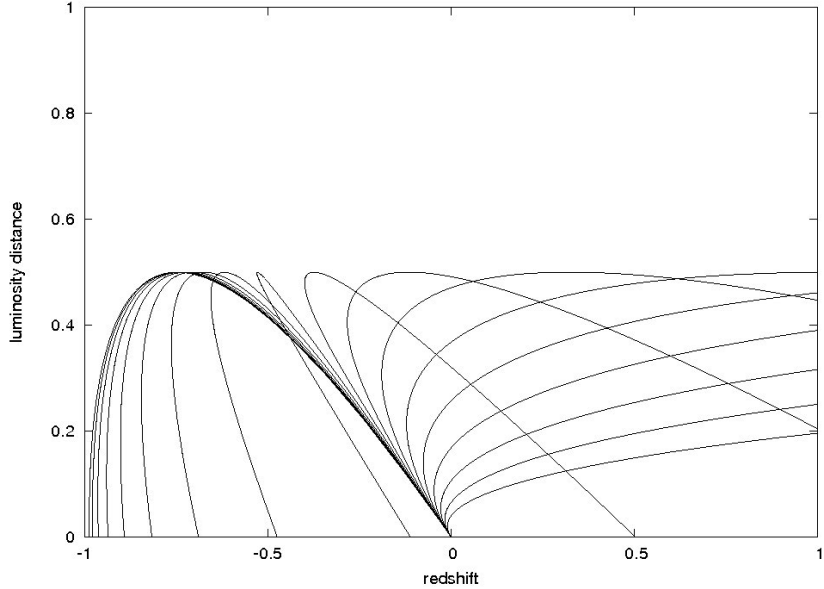
\sfig{-90}{0.45}{closedcollapsingFRW.jpg}{hubb-closedcollapsing}{The Hubble diagram in a closed, collapsing FRW universe. Different curves correspond to different $H_0a_0$ values. Note that the redshift can be negative (i.e., we may get a blueshift), and in that region the diagram is multivalued.}
We also need the expressions for light propagation through an arbitrary point in an arbitrary direction. Due to homogenity and isotropy, this can be obtained from the expressions above by performing a suitable rotation of the coordinate system (which is an isometry). This transformation may be represented as rotation of a three-sphere embedded into a four dimensional Eucledian space. Let the Cartesian coordinates of this space denoted by $x_1$, $x_2$, $x_3$ and $x_4$. The homogeneous and isotropic three dimensional space of our interest is the surface of a sphere of radius $a$. This surface may be parametrized as 
\eq{ 
x_1&=&a\sin\psi\sin\vartheta\cos\varphi\\ 
x_2&=&a\sin\psi\sin\vartheta\sin\varphi\\ 
x_3&=&a\sin\psi\cos\vartheta\\ 
x_4&=&a\cos\psi 
}{fc.k.7}  
 
Obviously, the usual radial coordinate $r$ is given by 
$$r=\sin\psi\;.$$ 
 
The line element on the surface is just the spatial metric we get from Eq.(\ref{eq:fc.00}), i.e. 
\eq{  
ds^2&=&a^2\left(\frac{d^2 r}{1-r^2}+r^2(d^2\vartheta+\sin^2\vartheta d^2\varphi)\right)\nonumber\\ 
&=&a^2\left(d^2\psi+\sin^2\psi(d^2\vartheta+\sin^2\vartheta d^2\varphi)\right) 
}{fc.k.8} 
 
A four-dimensional Eucledian rotation around any axis through the origin is an isometry of the three-surface. We shall use three such rotations in order to transfer the origin of the three-space (i.e., $r=0$) to another point whose coordinates are $\psi=\psi_k$, $\vartheta=\vartheta_k$ and $\varphi=\varphi_k$: 
\begin{enumerate} 
\item 
$x_3=const.\;, x_4=const.$ 
 
This readily implies that $r$ (or $\psi$) and $\vartheta$ are constants, too, while $\varphi$ changes according to 
\eq{ 
\left(\begin{array}{c}\cos\varphi'\\\sin\varphi'\end{array}\right)=\left(\begin{array}{cc}\cos\eta&-\sin\eta\\\sin\eta&\cos\eta\end{array}\right)\left(\begin{array}{c}\cos\varphi\\\sin\varphi\end{array}\right) 
}{fc.k.9} 
where $\eta$ stands for the angle of rotation. This implies 
$$\varphi'=\varphi+\eta\;,$$ 
hence choosing 
\eq{ 
\eta=-\varphi_k 
}{fc.k.10} 
 
sets the $\varphi$ coordinate of the new (primed) origin to $\varphi_k$.  
 
\item 
$x_2=const.\;, x_4=const.$ 
 
This readily implies that $r$ (or $\psi$) is constant, too, while $\vartheta$ and $\varphi$ change according to 
\eq{ 
\left(\begin{array}{c}\sin\vartheta'\cos\varphi'\\\cos\vartheta'\end{array}\right)=\left(\begin{array}{cc}\cos\gamma&-\sin\gamma\\\sin\gamma&\cos\gamma\end{array}\right)\left(\begin{array}{c}\sin\vartheta\cos\varphi\\\cos\vartheta\end{array}\right) 
}{fc.k.11} 
where $\gamma$ stands for the angle of rotation. This implies 
\eq{ 
\cos\vartheta'&=&\sin\gamma\sin\vartheta\cos\varphi+\cos\gamma\cos\vartheta\\ 
\cos\varphi'&=&\frac{\cos\gamma\sin\vartheta\cos\varphi-\sin\gamma\cos\vartheta}{\sqrt{1-\left(\sin\gamma\sin\vartheta\cos\varphi+\cos\gamma\cos\vartheta\right)^2}} 
}{fc.k.12} 
If $\varphi$ is already zero (due to the previous transformation), 
this implies 
$$\vartheta'=\vartheta-\gamma\;,\;\varphi'=0\;,$$ 
hence choosing 
\eq{ 
\gamma=\vartheta_k 
}{fc.k.13} 
 
sets the $\vartheta$ coordinate of the new (primed) origin to $\vartheta_k$ while keeping $\varphi$ unchanged.

\item 
$x_1=const.\;, x_2=const.$ 
 
This readily implies that $\varphi$ is constant, too, while $\vartheta$ and $\psi$ change according to 
\eq{ 
\left(\begin{array}{c}\sin\psi'\cos\vartheta'\\\cos\psi'\end{array}\right)=\left(\begin{array}{cc}\cos\nu&-\sin\nu\\\sin\nu&\cos\nu\end{array}\right)\left(\begin{array}{c}\sin\psi\cos\vartheta\\\cos\psi\end{array}\right) 
}{fc.k.14} 
where $\nu$ stands for the angle of rotation. This implies 
\eq{ 
\cos\psi'&=&\sin\nu\sin\psi\cos\vartheta+\cos\nu\cos\psi\\ 
\cos\vartheta'&=&\frac{\cos\nu\sin\psi\cos\vartheta-\sin\nu\cos\psi}{\sqrt{1-\left(\sin\nu\sin\psi\cos\vartheta+\cos\nu\cos\psi\right)^2}} 
}{fc.k.15} 
If $\vartheta$ is already zero (due to the previous transformation), 
this implies 
$$\psi'=\psi-\nu\;,\; \vartheta'=0\;,$$ 
hence choosing 
\eq{ 
\nu=\psi_k 
}{fc.k.16} 
 
sets the $\psi$ coordinate of the new (primed) origin to $\psi_k$ while keeping $\vartheta$ and $\varphi$ unchanged.

\end{enumerate} 
 
The succession of the above three rotations in the given order and with the given parameters corresponds to a spatial translation in the closed three-space.    
The resulting transformation may be expressed implicitly by 
\eq{ 
\sin\psi'\sin\vartheta'\cos\varphi'&=&\sin\psi\left(-\sin\vartheta_k\cos\vartheta+\cos\vartheta_k\sin\vartheta\cos(\varphi-\varphi_k)\right)\\ 
\sin\psi'\sin\vartheta'\sin\varphi'&=&\sin\psi\sin\vartheta\sin(\varphi-\varphi_k)\\ 
\sin\psi'\cos\vartheta'&=&\sin\psi\cos\psi_k\left(\cos\vartheta_k\cos\vartheta+\sin\vartheta_k\sin\vartheta\cos(\varphi-\varphi_k)\right)\nonumber\\ 
&&-\cos\psi\sin\psi_k\\ 
\cos\psi'&=&\sin\psi_k\sin\psi\left(\cos\vartheta_k\cos\vartheta+\sin\vartheta_k\sin\vartheta\cos(\varphi-\varphi_k)\right)\nonumber\\ 
&&+\cos\psi_k\cos\psi 
}{fc.k.17}

Similarly, the inverse transformation reads 
\eq{ 
\sin\psi\sin\vartheta\cos\varphi&=&\sin\vartheta_k\cos\varphi_k\left(\cos\psi_k\sin\psi'\cos\vartheta'+\sin\psi_k\cos\psi'\right)\nonumber\\ 
&&+\sin\psi'\sin\vartheta'\left(-\sin\varphi_k\sin\varphi'+\cos\vartheta_k\cos\varphi_k\cos\varphi'\right)\qquad\label{closed_t}\\ 
\sin\psi\sin\vartheta\sin\varphi&=&\sin\vartheta_k\sin\varphi_k\left(\cos\psi_k\sin\psi'\cos\vartheta'+\sin\psi_k\cos\psi'\right)\nonumber\\ 
&&+\sin\psi'\sin\vartheta'\left(\cos\varphi_k\sin\varphi'+\cos\vartheta_k\sin\varphi_k\cos\varphi'\right)\\ 
\sin\psi\cos\vartheta&=&\cos\vartheta_k\left(\sin\psi_k\cos\psi'+\cos\psi_k\sin\psi'\cos\vartheta'\right)\nonumber\\ 
&&-\sin\psi'\sin\vartheta'\cos\varphi'\sin\vartheta_k\\ 
\cos\psi&=&\cos\psi_k\cos\psi'-\sin\psi_k\sin\psi'\cos\vartheta' 
}{fc.k.18} 
 
These expressions describe the trajectory of the light when $\varphi'$ and
$\vartheta'$ are constants and
\eq{
\psi'&=&2(\chi-\chi_1)\\
t&=&\beta(\chi-\sin\chi\cos\chi)
}{fc.k.19}
i.e., when the light ray moves radially in the K' frame.

The expansion rate is
\eq{\Theta=\frac{\dot{a}}{a^2}+\frac{\sqrt{1-r'^2}}{a^2\;r'}=\frac{\dot{a}+\cot\psi'}{a^2}}{sw_c.3}
and for the cross section we have
\eq{A=\Omega a^2\;r'^2=\Omega a^2\;\sin^2\psi'\quad.}{sw_c.4}

Finally, we calculate the components of the four-velocity of the
light. We use the same technique, as in the case of the flat FRW universe.
In the K' frame we have
\eq{u^{t'}&=&\frac{1}{a}\\
u^{r'}&=&\frac{\cos\psi'}{a^2}\\
u^{\vartheta'}&=&0\\
u^{\varphi'}&=&0
}{fc.k.20}
Pulling down indices:
\eq{u_{t'}&=&\frac{1}{a}\\
u_{r'}&=&-\frac{1}{\cos\psi'}\\
u_{\vartheta'}&=&0\\
u_{\varphi'}&=&0
}{fc.k.21}
Applying Eq.(\ref{eq:xxx1}) we get in the K frame
\eq{
u_{t}&=&\frac{1}{a}\\
u_{r}&=&-\frac{\cos\psi_k\tan\psi-\sin\psi_k\left(\cos\vartheta_k\cos\vartheta+\sin\vartheta_k\sin\vartheta\cos(\varphi-\varphi_k)\right)}{\sin\psi'}\qquad\\
u_{\vartheta}&=&-\frac{\sin\psi_k\sin\psi\left(\cos\vartheta_k\sin\vartheta-\sin\vartheta_k\cos\vartheta\cos(\varphi-\varphi_k)\right)}{\sin\psi'}\\
u_{\varphi}&=&-\frac{\sin\psi_k\sin\psi\sin\vartheta_k\sin\vartheta\sin(\varphi-\varphi_k)}{\sin\psi'}
}{fc.k.22}
Finally, by pulling up the indices with the contravariant metric tensor we get
\eq{
u^{t}&=&\frac{1}{a}\\
u^{r}&=&\frac{\cos\psi}{a^2\sin\psi'}\Big[\cos\psi_k\sin\psi\Big.\nonumber\\
&&\Big.-\sin\psi_k\cos\psi\left(\cos\vartheta_k\cos\vartheta+\sin\vartheta_k\sin\vartheta\cos(\varphi-\varphi_k)\right)\Big]\label{closed_u_r}\\
u^{\vartheta}&=&\frac{\sin\psi_k\left(\cos\vartheta_k\sin\vartheta-\sin\vartheta_k\cos\vartheta\cos(\varphi-\varphi_k)\right)}{a^2\sin\psi'\sin\psi}\\
u^{\varphi}&=&\frac{\sin\psi_k\sin\vartheta_k\sin(\varphi-\varphi_k)}{a^2\sin\psi'\sin\psi\sin\vartheta}
}{fc.k.23}

\section{Light propagation in the Schwarzschild metric} 
The Schwarzschild  metric can be written as 
\eq{ds^2=\left(1-\frac{r_g}{r}\right)dt^2-\frac{dr^2}{1-\frac{r_g}{r}}-r^2\left(d\vartheta^2+\sin^2\vartheta d\varphi^2\right)\quad.}{fs.01} 
Hence the metric tensor is given by 
\eqq{g_{tt}&=&1-\frac{r_g}{r}}{fs.02} 
{g_{rr}&=&-\frac{1}{1-\frac{r_g}{r}}}{fs.03} 
{g_{\vartheta\vartheta}&=&-r^2}{fs.04} 
{g_{\varphi\varphi}&=&-r^2\sin^2 \vartheta}{fs.05} 
The non-vanishing components of the Christoffel symbol read 
\begin{eqnarray} 
\Gamma_{tr}^{t}&=&\Gamma_{rt}^{t}=-\Gamma_{rr}^{r}=\frac{r_g}{2r(r-r_g)}\label{eq:fs.06}\\ 
\Gamma_{tt}^{r}&=&\frac{r_g(r-r_g)}{2r^3}\label{eq:fs.07}\\ 
\Gamma_{\vartheta\vartheta}^{r}&=&r_g-r\label{eq:fs.08}\\ 
\Gamma_{\varphi\varphi}^{r}&=&(r_g-r)\sin^2 \vartheta\label{eq:fs.09}\\ 
\Gamma_{\varphi\varphi}^{\vartheta}&=&-\sin \vartheta \cos \vartheta\label{eq:fs.10}\\ 
\Gamma_{\vartheta r}^{\vartheta}&=&\Gamma_{r \vartheta}^{\vartheta}=\Gamma_{\varphi 
  r}^{\varphi}=\Gamma_{r \varphi}^{\varphi}=\frac{1}{r}\label{eq:fs.11}\\ 
\Gamma_{\vartheta\varphi}^{\varphi}&=&\Gamma_{\varphi\vartheta}^{\varphi}=\cot \vartheta\label{eq:fs.12} 
\end{eqnarray} 
Since we are dealing here with a vacuum solution of Einstein's equations, Ricci's tensor vanishes, hence Riemann's tensor is equal to Weyl's tensor. The nonzero components are 
\begin{eqnarray} 
R_{trtr}&=&C_{trtr}=\frac{r_g}{r^3}\label{eq:fs.18}\\ 
R_{t\vartheta t\vartheta}&=&C_{t\vartheta t\vartheta}=-\frac{r_g(r-r_g)}{2r^2}\label{eq:fs.19}\\ 
R_{t\varphi t\varphi}&=&C_{t\varphi t\varphi}=-\frac{r_g(r-r_g)\sin^2\vartheta}{2r^2}\label{eq:fs.20}\\ 
R_{r\vartheta r \vartheta}&=&C_{r\vartheta r \vartheta}=\frac{r_g}{2(r-r_g)}\label{eq:fs.21}\\ 
R_{r\varphi r\varphi}&=&C_{r\varphi r\varphi}=\frac{r_g\sin^2\vartheta}{2(r-r_g)}\label{eq:fs.22}\\ 
R_{\vartheta \varphi \vartheta \varphi}&=&C_{\vartheta \varphi \vartheta \varphi}=-r_gr\sin^2\vartheta\label{eq:fs.23} 
\end{eqnarray} 
where 
\begin{eqnarray} 
R_{ijkl}&=&-R_{jikl}=-R_{ijlk}\label{eq:fs.24}\\ 
R_{ijkl}&=&R_{klij}\label{eq:fs.25} 
\end{eqnarray} 
 
In a spherically symmetric field particle trajectories (including photons) are confined to a plane perpendicular to the conserved angular momentum vector. Provided that the angular momentum vector points towards the $z$-axis and the center of the field is at the origin, $\vartheta=\frac{\pi}{2}$ during the whole motion. The motion is fully integrable. Integration may be performed by solving the Hamilton-Jacobi equation which describes light propagation if $m=0$. Let us introduce the impact parameter $\rho=\frac{Jc}{E_0}$ instead of the angular momentum $J$. The orbit is defined by 
\eq{\varphi=\int{\frac{dr}{r^2\sqrt{\frac{1}{\rho^2}-\frac{1}{r^2}\left(1-\frac{r_g}{r}\right)}}}}{fs.26} 
which will be converted to an elliptic integral. To this end let us introduce 
\eq{\xi\equiv\frac{r_g}{r}\quad.}{fs.27} 
In terms of this new variable of integration we have  
\eq{\varphi=-\int{\frac{d\xi}{\sqrt{\xi^3-\xi^2+\frac{r_g^2}{\rho^2}}}}\quad.}{fs.28} 
The third order polynomial under the square root factorizes as 
\eq{\xi^3-\xi^2+\frac{r_g^2}{\rho^2}=(\xi-\xi_1)(\xi-\xi_2)(\xi-\xi_3)\quad,}{fs.29} 
where $\xi_1$, $\xi_2$ and $\xi_3$ are the roots of the polynomial, namely,  
\eqt{\xi_1&=&\frac{1}{3}\left(1+2\cos\left(\frac{\alpha+2\pi}{3}\right)\right)}{fs.33} 
{\xi_2&=&\frac{1}{3}\left(1+2\cos\left(\frac{\alpha-2\pi}{3}\right)\right)}{fs.34} 
{\xi_3&=&\frac{1}{3}\left(1+2\cos\left(\frac{\alpha}{3}\right)\right)}{fs.35} 
where 
\eq{\sin\left(\frac{\alpha}{2}\right)=\frac{3\sqrt{3}}{2}\frac{r_g}{\rho}\quad.}{fs.36} 
Now we can see that  
\eq{\xi_1<0<\xi_2<\xi_3<1\quad.}{fs.36a} 
Furthermore, during the motion $0\le\xi\le\xi_2$ holds and the turning point
of the light trajectory (i.e., its nearest point to the center of the field)
is at $\xi=\xi_2$. Note that for $\rho<3\sqrt{3}r_g/2$ no turning point exists
(cf. Eq.(\ref{eq:fs.36})), thus light beams with such a small impact parameter inevitably fall into the center of the field.  
 
If the angle is measured from the turning point,  
\eq{\varphi=-\int\limits_{\xi_2}^{\xi}{\frac{d\xi}{\sqrt{(\xi-\xi_1)(\xi-\xi_2)(\xi-\xi_3)}}}\;.}{fs.37} 
Performing the integration we have 
\eq{\varphi&=&\frac{\sqrt{2}\sqrt[4]{3}}{\sqrt{\sin\left(\frac{\alpha+\pi}{3}\right)}}\left[K\left(\sqrt{\frac{\sin\left(\frac{\alpha}{3}\right)}{\sin\left(\frac{\alpha+\pi}{3}\right)}}\right)\right.\nonumber\\ 
&&\left.-F\left(\sqrt{\frac{\sqrt{3}}{2}\frac{\xi-\xi_1}{\sin\left(\frac{\alpha}{3}\right)}},\;\sqrt{\frac{\sin\left(\frac{\alpha}{3}\right)}{\sin\left(\frac{\alpha+\pi}{3}\right)}}\right)\right]\quad.}{fs.38a} 
Especially, when the light comes from the infinity, the change of the azimuthal angle until reaching the turning point is  
\eq{\bar\varphi&=&\frac{\sqrt{2}\sqrt[4]{3}}{\sqrt{\sin\left(\frac{\alpha+\pi}{3}\right)}}\left[K\left(\sqrt{\frac{\sin\left(\frac{\alpha}{3}\right)}{\sin\left(\frac{\alpha+\pi}{3}\right)}}\right)\right.\nonumber\\ 
&&\left.-F\left(\frac{1}{\sqrt{2}}\sqrt{1-\frac{\sqrt{3}}{3}\tan{\frac{\alpha}{6}}},\;\sqrt{\frac{\sin\left(\frac{\alpha}{3}\right)}{\sin\left(\frac{\alpha+\pi}{3}\right)}}\right)\right]\quad.}{fs.38} 
In the above equations $F(z,k)$ stands for the incomplete elliptic integral of the first kind and $K(k)$ for the complete elliptic integral of the first kind, i.e. 
\eqd{K(k)&=&F(1,k)}{fs.39} 
{F(z,k)&=&\int\limits_{0}^{z}{\frac{dt}{\sqrt{(1-t^2)(1-k^2t^2)}}}}{fs.40} 
Gravitational field bends the light beam. Let us denote by $\delta\varphi$ the deviation of the outgoing (to infinity) light beam from the original (ingoing) direction. Explicitly, we have 
\eq{\delta\varphi=2\bar\varphi-\pi\quad.}{fs.41a} 
The series expansion of Eq.(\ref{eq:fs.38}) in terms of $r_g/\rho$ yields the classical expression for the deviation of light. The zeroth order vanishes, 
\eq{\delta\varphi^{(0)}=2\bar\varphi^{(0)}-\pi=0\quad,}{fs.41} 
and in first order we get 
\eq{\delta\varphi^{(1)}=\frac{2r_g}{\rho}\quad.}{fs.42} 
The time dependence $r=r(t)$ is determined by 
\eq{c\;t=\int{\frac{dr}{\left(1-\frac{r_g}{r}\right)\sqrt{1-\frac{\rho^2}{r^2}\left(1-\frac{r_g}{r}\right)}}}}{fs.43} 
or, using the new variable of integration introduced in Eq.(\ref{eq:fs.27})) 
\eq{-\frac{c\;t\;\rho}{r_g^2}=\int{\frac{d\xi}{\xi^2(1-\xi)\sqrt{(\xi-\xi_1)(\xi-\xi_2)(\xi-\xi_3)}}}\quad,}{fs.44} 
where the roots $\xi_1$, $\xi_2$ and $\xi_3$ are determined by Eqs.(\ref{eq:fs.33}--\ref{eq:fs.35}). Performing the integration we get 
\eq{t&=&\pm\frac{\rho}{c}\frac{\sqrt[4]{3}}{\sqrt{2}\sqrt{\sin\left(\frac{\alpha+\pi}{3}\right)}}\left[\frac{\sqrt{2}}{\sqrt[4]{3}}\sqrt{\sin\left(\frac{\alpha+\pi}{3}\right)}\;\frac{1}{\xi}\sqrt{\xi^3-\xi^2+\frac{r_g^2}{\rho^2}}\right.\nonumber\\ 
&&+\frac{1}{3}\left(1+2\cos\left(\frac{\alpha+2\pi}{3}\right)\right)F\left(\sqrt{\frac{\sqrt{3}}{2}\frac{\xi-\xi_1}{\sin\left(\frac{\alpha}{3}\right)}},\;\sqrt{\frac{\sin\left(\frac{\alpha}{3}\right)}{\sin\left(\frac{\alpha+\pi}{3}\right)}}\right)\nonumber\\ 
&&+\frac{2}{\sqrt{3}}\sin\left(\frac{\alpha+\pi}{3}\right)E\left(\sqrt{\frac{\sqrt{3}}{2}\frac{\xi-\xi_1}{\sin\left(\frac{\alpha}{3}\right)}},\;\sqrt{\frac{\sin\left(\frac{\alpha}{3}\right)}{\sin\left(\frac{\alpha+\pi}{3}\right)}}\right)\nonumber\\ 
&&-\frac{r_g^2}{\rho^2}\frac{6}{1+2\cos\left(\frac{\alpha+2\pi}{3}\right)}\Pi\left(\sqrt{\frac{\sqrt{3}}{2}\frac{\xi-\xi_1}{\sin\left(\frac{\alpha}{3}\right)}},\frac{-2\sqrt{3}\sin\left(\frac{\alpha}{3}\right)}{1+2\cos\left(\frac{\alpha+2\pi}{3}\right)},\sqrt{\frac{\sin\left(\frac{\alpha}{3}\right)}{\sin\left(\frac{\alpha+\pi}{3}\right)}}\right)\nonumber\\ 
&&+\frac{r_g^2}{\rho^2}\frac{3}{1-\cos\left(\frac{\alpha+2\pi}{3}\right)}\Pi\left(\sqrt{\frac{\sqrt{3}}{2}\frac{\xi-\xi_1}{\sin\left(\frac{\alpha}{3}\right)}},\frac{\sqrt{3}\sin\left(\frac{\alpha}{3}\right)}{1-\cos\left(\frac{\alpha+2\pi}{3}\right)},\sqrt{\frac{\sin\left(\frac{\alpha}{3}\right)}{\sin\left(\frac{\alpha+\pi}{3}\right)}}\right)\nonumber\\ 
\Bigg.&&-C\Bigg]\quad.}{fs.45} 
Here the $+$ sign holds if the light is going away and the $-$ sign, if it is approaching. The symbol $C$ is a constant of integration. We choose it so that $t=0$ corresponds to the turning point, i.e., $\xi=\xi_2$. Hence 
\eq{C&=&\frac{1}{3}\left(1+2\cos\left(\frac{\alpha+2\pi}{3}\right)\right)K\left(\sqrt{\frac{\sin\left(\frac{\alpha}{3}\right)}{\sin\left(\frac{\alpha+\pi}{3}\right)}}\right)\nonumber\\ 
&&+\frac{2}{\sqrt{3}}\sin\left(\frac{\alpha+\pi}{3}\right)E\left(\sqrt{\frac{\sin\left(\frac{\alpha}{3}\right)}{\sin\left(\frac{\alpha+\pi}{3}\right)}}\right)\nonumber\\ 
&&-\frac{r_g^2}{\rho^2}\frac{6}{1+2\cos\left(\frac{\alpha+2\pi}{3}\right)}\Pi\left(\frac{-2\sqrt{3}\sin\left(\frac{\alpha}{3}\right)}{1+2\cos\left(\frac{\alpha+2\pi}{3}\right)},\sqrt{\frac{\sin\left(\frac{\alpha}{3}\right)}{\sin\left(\frac{\alpha+\pi}{3}\right)}}\right)\nonumber\\ 
&&+\frac{r_g^2}{\rho^2}\frac{3}{1-\cos\left(\frac{\alpha+2\pi}{3}\right)}\Pi\left(\frac{\sqrt{3}\sin\left(\frac{\alpha}{3}\right)}{1-\cos\left(\frac{\alpha+2\pi}{3}\right)},\sqrt{\frac{\sin\left(\frac{\alpha}{3}\right)}{\sin\left(\frac{\alpha+\pi}{3}\right)}}\right)\quad.}{fs.45a} 
In Eq.(\ref{eq:fs.45}) $E(z,k)$ stands for the incomplete elliptic integral of the second kind and $\Pi(z,\nu,k)$ for the incomplete elliptic integral of the third kind. I.e.,  
\eqd{E(z,k)&=&\int\limits_{0}^{z}{\sqrt{\frac{1-k^2t^2}{1-t^2}}dt}}{fs.46} 
{\Pi(z,\nu,k)&=&\int\limits_{0}^{z}{\frac{dt}{\left(1-\nu t^2\right)\sqrt{(1-t^2)(1-k^2t^2)}}}}{fs.47} 
In Eq.(\ref{eq:fs.45a}) $E(k)$ stands for the complete elliptic integral of the second kind and $\Pi(\nu,k)$ for the complete elliptic integral of the third kind, i.e.,  
\eqd{E(k)&=&E(1,k)=\int\limits_{0}^{1}{\sqrt{\frac{1-k^2t^2}{1-t^2}}dt}}{fs.46a} 
{\Pi(\nu,k)&=&\Pi(1,\nu,k)=\int\limits_{0}^{1}{\frac{dt}{\left(1-\nu t^2\right)\sqrt{(1-t^2)(1-k^2t^2)}}}}{fs.47a} 
 
Eqs. (\ref{eq:fs.26}), (\ref{eq:fs.43}) imply 
\eqd{\frac{u^r}{u^{\varphi}}&=&\frac{dr}{d\varphi}=r^2\sqrt{\frac{1}{\rho^2}-\frac{1}{r^2}\left(1-\frac{r_g}{r}\right)}}{fs.48} 
{\frac{u^r}{u^t}&=&\frac{dr}{dt}=\left(1-\frac{r_g}{r}\right)\sqrt{1-\frac{\rho^2}{r^2}\left(1-\frac{r_g}{r}\right)}}{fs.49} 
Further, since $\vartheta=\frac{\pi}{2}$, we have 
\eq{u^{\vartheta}=0\;.}{fs.50} 
Therefore, the $r$ component of the geodetical equation sounds 
\eq{\frac{du^r}{dr}u^r+\Gamma^r_{tt}(u^t)^2+\Gamma^r_{rr}(u^r)^2+\Gamma^r_{\varphi\varphi}(u^{\varphi})^2=0\quad,}{fs.51} 
which implies according to Eqs.(\ref{eq:fs.06} -- \ref{eq:fs.09}) 
\eq{\frac{du^r}{dr}=\frac{1}{u^r}\frac{r_g}{2r^2}(1-\frac{r_g}{r})\left[\frac{1}{(1-\frac{r_g}{r})^2}(u^r)^2-(u^t)^2+\frac{2r^3}{r_g}(u^{\varphi})^2\right]\quad.}{fs.52} 
Using Eqs.(\ref{eq:fs.48}, \ref{eq:fs.49}) we get 
\eq{\frac{du^r}{dr}=\frac{r-\frac{3}{2}r_g}{\frac{r^4}{\rho^2}\left(1-\frac{\rho^2}{r^2}\left(1-\frac{r_g}{r}\right)\right)}u^r}{fs.53} 
with the solution 
\eq{u^r&=&u_0\exp{\left(\int{\frac{r-\frac{3}{2}r_g}{\frac{r^4}{\rho^2}\left(1-\frac{\rho^2}{r^2}\left(1-\frac{r_g}{r}\right)\right)}dr}\right)}\nonumber\\ 
&=&u_0\;\sqrt{1-\frac{\rho^2}{r^2}\left(1-\frac{r_g}{r}\right)}\quad,}{fs.54} 
where $u_0$ is an integration constant. Consequently, the four-velocity of the light in Schwarzschild metric is given by 
\eqq{u^{t}&=&u_0\;\frac{1}{1-\frac{r_g}{r}}}{fs.13} 
{u^{r}&=&u_0\;\sqrt{1-\frac{\rho^2}{r^2}\left(1-\frac{r_g}{r}\right)}}{fs.14} 
{u^{\vartheta}&=&0}{fs.15} 
{u^{\varphi}&=&u_0\;\frac{\rho}{r^2}}{fs.16} 
In order to determine the light intensity during propagation one has to calculate the change of the cross section of the light beam. Suppose that the centerline of the narrow beam of an elliptic cross section lies in the $\vartheta=\frac{\pi}{2}$ plane. This plane is clearly a symmetry plane of the beam, hence one of the axes of the cross section ellipse remains collateral with this plane and the other is perpendicular to it (assuming initial conditions already having possessed this property). We have to consider two type of trajectories on the outer surface of the beam: 
\begin{itemize} 
\item trajectories lying in the $\vartheta=\frac{\pi}{2}$ plane (type 1) 
\item trajectories at a maximal distance from the $\vartheta=\frac{\pi}{2}$ plane (type 2)
\end{itemize}
Certainly we have no further linearly independent possibilities for the deviations. 
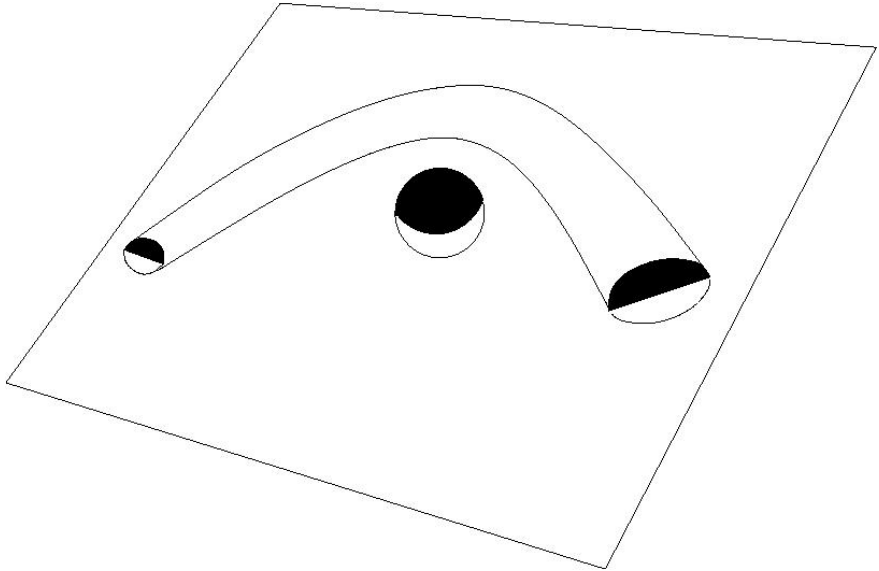
\sfig{0}{.45}{beam1.jpg}{beam1}{Light beam near a spherical massive object. The plane shown is that of the centerline of the beam.}
\newpage
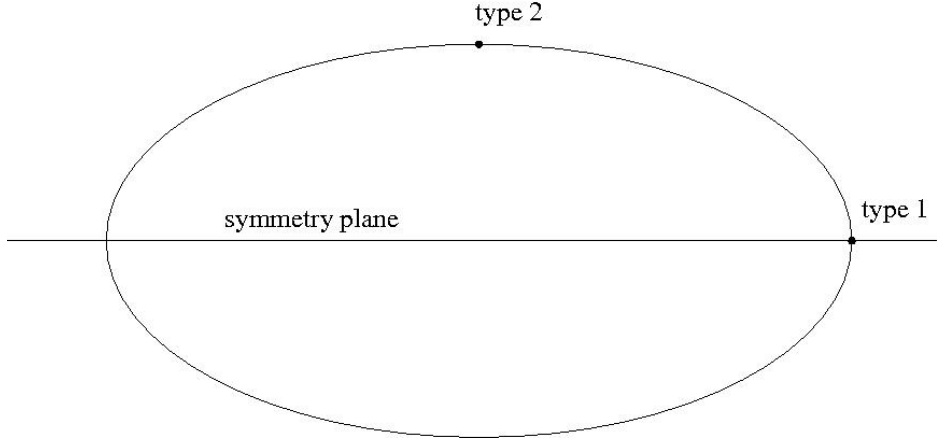
\sfig{0}{.45}{beam2.jpg}{beam2}{Cross section of the beam. The symmetry plane and orbits type 1 and 2 are shown.}

Trajectories of type 1 lie in the same plane as the centerline, but their initial conditions and impact parameter slightly deviate from those of the reference centerline trajectory. The perpendicular distance $dl$ from the centerline may be expressed as 
\eq{dl=r\sqrt{1-\frac{\rho^2}{r^2}\left(1-\frac{r_g}{r}\right)}\left(d\varphi-\frac{\partial \varphi}{\partial r}dr\right) }{xx.1} 
Similarly, at the initial point (we consider $\varphi$ to be the function of $r$, $\rho$ and the initial position $r_1$ and $\varphi_1$) we have 
\eq{dl_1=r_1\sqrt{1-\frac{\rho^2}{r_1^2}\left(1-\frac{r_g}{r_1}\right)}\left(\frac{\partial \varphi}{\partial \varphi_1}d\varphi_1+\frac{\partial \varphi}{\partial r_1}dr_1\right) }{xx.2} 
Also, we need to take into account that the total change in the value of the angle may be written as 
 \eq{d\varphi= \frac{\partial \varphi}{\partial r}dr+\frac{\partial \varphi}{\partial \rho}d\rho+\frac{\partial \varphi}{\partial \varphi_1}d\varphi_1+\frac{\partial \varphi}{\partial r_1}dr_1}{xx.3} 
Combining Eqs.(\ref{eq:xx.1}), (\ref{eq:xx.2}), (\ref{eq:xx.3}), (\ref{eq:fs.26}) we get for the expansion rate for trajectories in the plane $\vartheta=\frac{\pi}{2}$ 
\eq{\frac{dl}{dl_1}&=&r\sqrt{1-\frac{\rho^2}{r^2}\left(1-\frac{r_g}{r}\right)}\left[\frac{d\rho}{dl_1}\int\limits_{r_1}^r{\frac{dr}{r^2\left(1-\frac{\rho^2}{r^2}\left(1-\frac{r_g}{r}\right)\right)^{\frac{3}{2}}}}\right.\nonumber\\ 
&&\left.+\frac{1}{r_1\sqrt{1-\frac{\rho^2}{r_1^2}\left(1-\frac{r_g}{r_1}\right)}}\right] }{xx.4} 
As for trajectories type 2, they also lie in some plane which makes some small angle $d\gamma$  with the original $\vartheta=\frac{\pi}{2}$ plane. 
Simple geometrical considerations imply that the perpendicular distance from the reference trajectory may be written as 
\eq{r\left|\sin(\varphi-\Phi)\right|d\gamma}{xx.5} 
where $\Phi$ denotes the azimuthal angle of the intersection line of the planes. 
Hence the expansion rate for trajectories not in the plane $\vartheta=\frac{\pi}{2}$ we get 
\eq{\frac{r\left|\sin(\varphi-\Phi)\right|}{r_1\left|\sin(\varphi_1-\Phi)\right|}}{xx.6} 
Note that eventually we let the initial point coincide with the light source that implies $\Phi=\varphi_1$. A direct substitution at this stage is obviously not yet allowed. 
  
Putting things together, 
if the light beam has cross section $A_1$ at the point $(r_1,\varphi_1)$, then  at another point $(r,\varphi)$ its cross section is 
\eq{A&=&\frac{r^2\left|\sin(\varphi-\Phi)\right|}{r_1\left|\sin(\varphi_1-\Phi)\right|}\sqrt{1-\frac{\rho^2}{r^2}\left(1-\frac{r_g}{r}\right)}\left[\frac{d\rho}{dl_1}\int\limits_{r_1}^r{\frac{dr}{r^2\left(1-\frac{\rho^2}{r^2}\left(1-\frac{r_g}{r}\right)\right)^{\frac{3}{2}}}}\right.\nonumber\\ 
&&\left.+\frac{1}{r_1\sqrt{1-\frac{\rho^2}{r_1^2}\left(1-\frac{r_g}{r_1}\right)}}\right]A_1}{fs.A1}

Below we cast this equation to a more suitable form. 
Let us denote by $\mu_1$ the angle of the light ray and the radial direction at the initial point $(r_1,\varphi_1)$ in the locally Eucledian frame. 

Geometry implies 
\eqd{\cos\mu_1&\propto&\sqrt{-g_{rr}}dr=\frac{dr_1}{\sqrt{1-\frac{r_g}{r_1}}}}{fs.k.12} 
{\sin\mu_1&\propto&r_1d\varphi_1\;,}{fs.k.13} 
the proportionality constant being the same in both equations.
Using Eq.(\ref{eq:fs.26}) we obtain 
\eq{\cot\mu_1=\frac{dr_1}{d\varphi_1}\frac{1}{r_1\sqrt{1-\frac{r_g}{r_1}}}=\frac{r_1}{\sqrt{1-\frac{r_g}{r_1}}}\sqrt{\frac{1}{\rho^2}-\frac{1}{r_1^2}\left(1-\frac{r_g}{r_1}\right)}\quad,}{fs.k.14} 
therefore 
\eq{\sin\mu_1&=&\frac{\rho}{r_1}\sqrt{1-\frac{r_g}{r_1}}}{fs.k.01} 
Provided that the initial point is at a small distance $L$ from the source we may express the aperture $d\mu_1\ll 1$ of the light as
 \eq{d\mu_1=\frac{dl_1}{L}}{fs.k.15}
Eqs.(\ref{eq:fs.k.01}) and (\ref{eq:fs.k.15}) readily imply that
\eq{\frac{d\rho}{dl_1}=\frac{r_1\cos\mu_1}{L\sqrt{1-\frac{r_g}{r_1}}}}{fs.k.16}
Suppose for simplicity that the light beam has a circular cross section near its source. Then the distance (\ref{eq:xx.5}) may also be expressed as
\eq{r_1\left|\sin(\varphi_1-\Phi)\right|d\gamma=L\sin\mu_1d\gamma\quad.}{fs.k.02}  
Near the source, when $L\to0$, in Eq.(\ref{eq:fs.A1}) the term containing  
$d\rho/dl_1$ 
 becomes dominant, so the other term may be neglected. 
 Introducing the solid angle
\eq{\Omega=\frac{A_1}{L^2}\quad,}{fs.k.04} 
Eqs.(\ref{eq:fs.A1}), (\ref{eq:fs.k.16}) and (\ref{eq:fs.k.02}) lead to 
\eq{A&=&\frac{r_1^2r_0^2\left|\sin(\varphi_0-\varphi_1)\right|\Omega\sqrt{1-\frac{\rho^2}{r_0^2}\left(1-\frac{r_g}{r_0}\right)}\sqrt{1-\frac{\rho^2}{r_1^2}\left(1-\frac{r_g}{r_1}\right)}}{\rho\left(1-\frac{r_g}{r_1}\right)}\nonumber\\
&&\times\int\limits_{r_1}^{r_0}{\frac{dr}{r^2\left(1-\frac{\rho^2}{r^2}\left(1-\frac{r_g}{r}\right)\right)^{\frac{3}{2}}}}\qquad}{fs.A2} 
Note that at this stage we have been allowed to let the initial point coincide with the source, hence $\Phi=\varphi_1$ has been substituted. Also, the final point is denoted by ($r_0$, $\varphi_0$) henceforth.
Using the equation of the path (Eq.(\ref{eq:fs.38a})) we have
\eq{\varphi_0-\varphi_1&=&\frac{\sqrt{2}\sqrt[4]{3}}{\sqrt{\sin\left((\alpha+\pi)/3\right)}}\left[F\left(\sqrt{\frac{\sqrt{3}}{2}\frac{r_g/r_1-\xi_1}{\sin\left(\alpha/3\right)}},\;\sqrt{\frac{\sin\left(\alpha/3\right)}{\sin\left((\alpha+\pi)/3\right)}}\right)\right.\nonumber\\
&&-\left.F\left(\sqrt{\frac{\sqrt{3}}{2}\frac{r_g/r_0-\xi_1}{\sin\left(\alpha/3\right)}},\;\sqrt{\frac{\sin\left(\alpha/3\right)}{\sin\left((\alpha+\pi)/3\right)}}\right)
\right]\quad.}{fs.A3}
The redshift is given by 
\eq{z=\frac{\sqrt{g_{tt}(t_1)}u^t(t_1)}{\sqrt{g_{tt}(t_0)}u^t(t_0)}-1\quad,}{fs.k.05} 
hence inserting $u^t$ from Eq.(\ref{eq:fs.13}) we get 
\eq{z=\frac{\sqrt{1-\frac{r_g}{r_0}}}{\sqrt{1-\frac{r_g}{r_1}}}-1\quad.}{fs.k.06} 
Solving this for $r_1$ we have 
\eq{r_1=\frac{r_g(1+z)^2}{(1+z)^2-\left(1-\frac{r_g}{r_0}\right)}\quad.}{fs.k.08} 
The luminosity distance is 
\eq{d_L=(1+z)\sqrt{\frac{A}{\Omega}}\quad,}{fs.k.07} 
which, combining with Eqs.(\ref{eq:fs.A2}), (\ref{eq:fs.A3}) and (\ref{eq:fs.k.08}) yields the Hubble diagram. 
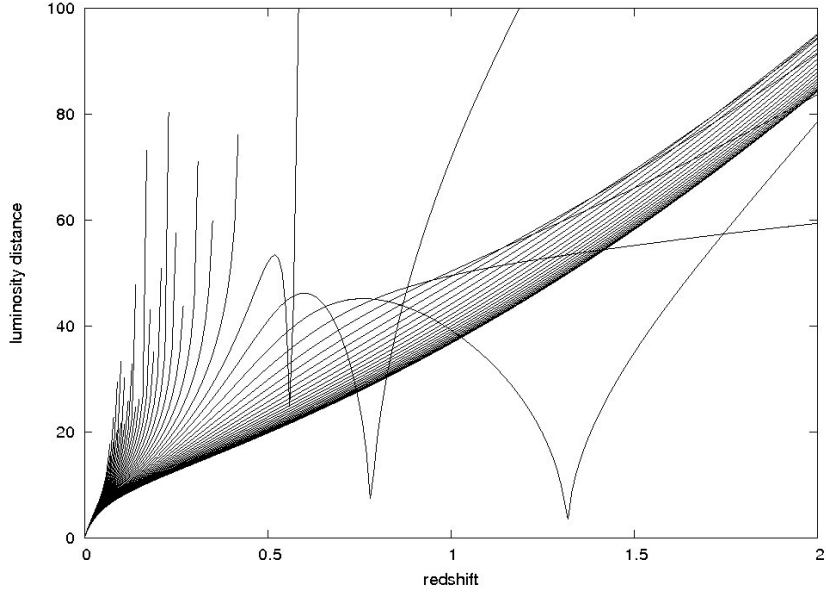
\sfig{-90}{0.45}{hubble_sch.jpg}{hubb-sch}{The Hubble-diagram in Schwarzschild-metric. The luminosity distance is given in $r_g$ units.}\\ 
In case of a different $\rho$ we get a different Hubble-diagram. In Fig.(\ref{fig:hubb-sch}) the lower envelope is determined by $\rho=0$, which corresponds to radial light propagation. In that case we have 
\eq{\varphi&=&\rho\left(\frac{1}{r_1}-\frac{1}{r}\right)\quad,}{fs.k.10a} 
which implies according to Eq.(\ref{eq:fs.A2}) 
\eq{A=\frac{\Omega(r_0-r_1)^2}{1-\frac{r_g}{r_1}}\quad,}{fs.k.10} 
which, together with Eqs.(\ref{eq:fs.k.08}), (\ref{eq:fs.k.07}) yields the luminosity distance 
\eq{d_L=r_0\sqrt{1-\frac{r_g}{r_0}}\frac{z(2+z)(1+z)^2}{(1+z)^2-\left(1-\frac{r_g}{r_0}\right)}\quad.}{fs.k.11} 
Concluding this section, we calculate the expansion rate $\Theta$ by using Eqs.(\ref{he1.20}), (\ref{eq:fs.14}), (\ref{eq:fs.16}) and (\ref{eq:fs.A2}). We get
\eq{
\Theta&=&\frac{1}{2A}\frac{dA}{d\lambda}=\frac{1}{2A}\left(\frac{\partial A}{\partial r}u^r+\frac{\partial A}{\partial \varphi}u^\varphi
\right)\nonumber\\
&=&u_0\left\{\frac{\rho}{2r^2}\cot(\varphi-\varphi_1)
+\frac{1-\frac{\rho^2}{2r}\left(1-\frac{r_g}{2r}\right)}{\sqrt{1-\frac{\rho^2}{r}\left(1-\frac{r_g}{r}\right)}}
\right.\nonumber\\
&&+\left.
\left[2\left(1-\frac{\rho^2}{r}\left(1-\frac{r_g}{r}\right)\right)\int_{r_1}^r\frac{dr}{r^2}\left(1-\frac{\rho^2}{r}\left(1-\frac{r_g}{r}\right)\right)^{-\frac{3}{2}}\right]^{-1}
\right\}
\qquad}{exp_rate_s}
\section{Matching conditions} 
Matching the different regions at the bordering spherical surfaces can be done explicitly in terms of suitable coordinate transformations. This transformation
is specified - to the necessary degree - by the requirement of continuity of the metric and its first derivatives when crossing the bordering surface.
Below we list the expressions obtained. For completeness, derivation is given in the Appendices. 

\subsection{Matching the flat FRW metric to the Schwarzschield metric} 
Metric outside of the sphere (flat FRW metric): 
\begin{eqnarray} 
ds^2=dt_f^2-a\left(t_f\right)^2\left(dr_f^2 
+r_f^2\left(d\vartheta^2+\sin^2\vartheta d\varphi^2\right)\right) 
\end{eqnarray} 
Metric inside of the sphere (Schwarzschield metric): 
\begin{eqnarray} 
ds^2=\left(1-\frac{r_g}{r_s}\right)dt_s^2-\frac{dr_s^2}{1-\frac{r_g}{r_s}} 
-r_s^2\left(d\vartheta^2+\sin^2\vartheta d\varphi^2\right)
\end{eqnarray} 
The bordering spherical surface lies at $r_f=R_1$. 
 
Coordinate transformation: 
\begin{eqnarray} 
t_s&=&t_f+3 \frac{r_g}{a_0R_1}\left(\frac{t_f}{t_0}\right)^{\frac{1}{3}}-2r_g\arth\left(\left(\frac{t_f}{t_0}\right)^{\frac{1}{3}}\sqrt{\frac{a_0R_1}{r_g}}\right)\nonumber\\
&&+(r_f-R_1)\frac{6a_0^2R_1t_f}{9t_0^{\frac{4}{3}}t_f^\frac{2}{3}-4a_0^2R_1^2}+(r_f-R_1)^2\frac{3a_0^2t_f\left(9t_0^{\frac{4}{3}}t_f^{\frac{2}{3}}-8a_0^2R_1^2\right)}{\left(9t_0^{\frac{4}{3}}t_f^{\frac{2}{3}}-4a_0^2R_1^2\right)^2}\qquad\label{match_fs_t}\\ 
r_s&=&\left(\frac{t_f}{t_0}\right)^{\frac{2}{3}}a_0r_f+(r_f-R_1)^2\frac{a_0^3R_1}{3t_0^2}\label{match_fs_r}\\
\vartheta_s&=&\vartheta_f\label{match_fs_theta}\\
\varphi_s&=&\varphi_f\label{match_fs_phi}
\end{eqnarray}
The matching condition is the following:
\begin{eqnarray} 
a_0^3R_1^3H_0^2=r_g 
\end{eqnarray} 

\subsection{Matching closed FRW metric to the Schwarzschield metric} 
Metric outside of the sphere (Schwarzschield metric): 
\begin{eqnarray}
ds^2=\left(1-\frac{r_g}{r_s}\right)dt_s^2-\frac{dr_s^2}{1-\frac{r_g}{r_s}} 
-r_s^2\left(d\vartheta^2+\sin^2\vartheta d\varphi^2\right)
\end{eqnarray} 
Metric inside of the sphere (closed FRW metric): 
\begin{eqnarray}
ds^2=dt_c^2-a\left(t_c\right)^2\left(\frac{dr_c^2}{1-r_c^2}+r_c^2\left(d\vartheta^2+\sin^2\vartheta d\varphi^2\right)\right)
\end{eqnarray} 
The bordering spherical surface lies at $r_c=R_2$.

In this case $a(t_c)$ fulfills Friedmann's equation whose parametric solution is given in Eqs.(\ref{eq:fc.00a})-(\ref{eq:fc.0xd}).
 
It is convenient to express the transformation laws in terms of this parametrization.  
Introducing the constant angle $\zeta$ by  
\begin{eqnarray} 
R_2=\sin\zeta 
\end{eqnarray} 
we get 
\begin{eqnarray} 
t_s&=&\beta\left[\cos\zeta(1+2\sin^2\zeta)\chi-\cos\zeta\sin\chi\cos\chi-2\sin^3\zeta\arth\left(\frac{\tan\chi}{\tan\zeta }\right)\right.\nonumber\\ 
&&+\left.(r_c-\sin\zeta)\frac{\tan\zeta\sin^3\chi\cos\chi}{\cos^2\zeta-\cos^2\chi}\right.\nonumber\\ 
&&+\left.\frac{1}{2}(r_c-\sin\zeta)^2\frac{\sin^3\chi\cos\chi\left(\cos^4\zeta-\cos^2\chi\right)}{\cos^3\zeta\left(\cos^2\zeta-\cos^2\chi\right)^2}\right]\label{match_cs_t}\\ 
r_s&=&\beta\left[r_c\sin^2\chi+\frac{3}{4}(r_c-\sin\zeta)^2\frac{\sin\zeta}{\cos^2\zeta}\right]\label{match_cs_r}\\
\vartheta_s&=&\vartheta_c\label{match_cs_theta}\\
\varphi_s&=&\varphi_c\label{match_cs_phi}
\end{eqnarray} 
Further, the consistency condition 
\begin{eqnarray} 
R_2^3\beta=r_g 
\end{eqnarray} 
must be fulfilled. The constant $\beta$ is defined by Eq.(\ref{eq:fc.0xb}).
\section{Light propagation and beam expansion in the Swiss-cheese model} 

Now we put everything together: using the previous analytic expressions we describe how to determine redshift and light intensity if the light ray passes through several regions, each having one of the three metrics considered. Above we considered these regions separately, therefore, we have expressions for light trajectories, cross sections and expansion rates when the light source is in the same region. 

The next problem is what happens at the borders. There Eqs.(\ref{match_fs_t})-(\ref{match_fs_phi}) or Eqs.(\ref{match_cs_t})-(\ref{match_cs_phi}), respectively, describe the coordinate transformation between the regions preserving the continuity of the metric and its first derivatives at the border. The four velocity transforms accordingly, namely,
\eq{u^t_s&=&\frac{u^t_f}{1-\frac{r_g}{a_0R_1}\left(\frac{t_f}{t_0}\right)^{-\frac{2}{3}}}+\frac{\frac{2}{3}\frac{a_0^2R_1}{t_0}\left(\frac{t_f}{t_0}\right)^{\frac{1}{3}}\;u^r_f}{1-\frac{r_g}{a_0R_1}\left(\frac{t_f}{t_0}\right)^{-\frac{2}{3}}}\\
u^r_s&=&\frac{2}{3}a_0r_ft_0^{-\frac{2}{3}}t_f^{-\frac{1}{3}}u^t_f+a_0\left(\frac{t_f}{t_0}\right)^{\frac{2}{3}}u^r_f\\
u^\vartheta_s&=&u^\vartheta_f\\
u^\varphi_s&=&u^\varphi_f
}{sw_fs.1}
at the flat FRW-Schwarzschild border and
\eq{u_s^t&=&u_c^t\frac{\cos\zeta\sin^2\chi}{\sin^2\chi-\sin^2\zeta}+
u_c^r\beta\frac{\tan\zeta\sin^3\chi\cos\chi}{\sin^2\chi-\sin^2\zeta}\\
u_s^r&=&u_c^tr_c\cot\chi +u_c^r\beta\sin^2\chi \\
u^\vartheta_s&=&u^\vartheta_c\\
u^\varphi_s&=&u^\varphi_c}{sw_cs.1}
at the closed FRW-Schwarzschild border.
The expansion rate and the cross section are invariant at the border since they are directly measurable quantities. Position, velocity, expansion rate and cross section render possible the calculation of these same quantities beyond the border, inside the other region at a later time. Below we describe how this can be achieved.

If the light ray enters the flat FRW region at time $t=t_{in}$ then Eq.(\ref{eq:sw_f.3}) determines $r'$ at this instant of time, given the expansion rate $\Theta$. The quantity $r'$ obtained would be the coordinate distance from the source had the light ray travelled all the time in flat FRW metric, like in a flat FRW universe. Next, using Eq.(\ref{tr2}) with $t=t_{in}$ determines $t_1$, which would be the emission time from the source in a flat FRW universe. Given the cross section $A$ at the entrance time, Eq.(\ref{eq:sw_f.4}) determines the solid angle $\Omega$ of the light ray at the imaginary source. 
The light trajectory is given by Eqs. (\ref{tr1a})-(\ref{tr1}) where the parameters $r_k$, $\vartheta_k$, $\varphi_k$, $\vartheta'$ and $\varphi'$ are determined from the position and velocity at $t=t_{in}$, using Eqs. (\ref{tr1a})-(\ref{tr1})  and Eqs. (\ref{flat_ur})-(\ref{flat_uphi}) together, where $r'$ is the value determined from the expansion rate as discussed above. Note that while there seems to be more equations than variables, in the reality  Eqs. (\ref{flat_ur})-(\ref{flat_uphi}) are not independent, since they satisfy 
$$a^2\left(\left(u^r\right)^2+r^2\left(u^\vartheta\right)^2+r^2\sin^2\vartheta\left(u^\varphi\right)^2\right)=\left(u^t\right)^2\quad.$$

If the light ray enters a closed FRW region the procedure is analogous to that used in the flat case. Given the expansion rate $\Theta$, at entrance time Eq.(\ref{eq:sw_c.3}) determines $r'$, or, equivalently, $\psi'$, the coordinate distance of the imaginary source (which would be there if light had travelled all the time in a closed FRW universe). Next, Eq.(\ref{eq:fc.k.19}) determines the emission time $t_1$, while Eq.(\ref{eq:sw_c.4}) yields the solid angle $\Omega$ of the light ray at the imaginary source. The light trajectory is given by Eqs. (\ref{closed_t})-(\ref{eq:fc.k.19}) where the parameters $\psi_k$, $\vartheta_k$, $\varphi_k$, $\vartheta'$ and $\varphi'$ are determined from the position and velocity at entrance time, using Eqs. (\ref{closed_t})-(\ref{eq:fc.k.18})  and Eqs. (\ref{closed_u_r})-(\ref{eq:fc.k.23}) together.

Finally, when the light ray enters a Schwarzschild region, Eqs.(\ref{eq:exp_rate_s}) and (\ref{eq:fs.A3}) determine $r_1$ and $\varphi_1$ (for the given the expansion rate $\Theta$), i.e., the position of the imaginary light source. The emission time $t_1$ may be found from Eq.(\ref{eq:fs.45}). Eq.(\ref{eq:fs.A2}) determines the solid angle $\Omega$ of the light ray near the source. Eqs.(\ref{eq:fs.38a}), (\ref{eq:fs.45}) describe the path and the time dependence, while Eqs.(\ref{eq:exp_rate_s}), (\ref{eq:fs.A2}) yield the expansion rate and the cross section at any later time.  

If there are several spheres (``holes in the Swiss-cheese''),   
the above procedures should be applied in succession. Since the matching formulae assume that the origins of the different coordinate systems coincide, shifting from one sphere to the other necessitates a spatial translation in the flat FRW region, given by Eqs.(\ref{tr0a})-(\ref{tr0}) or (\ref{tr1a})-(\ref{tr1}). Here $r_k$, $\vartheta_k$ and $\varphi_k$ are the coordinates of the centre of the next sphere in the old (unprimed) coordinate system. The velocity components transform accordingly. Further, a rotation around an axis through the origin may also be necessary afterwards, in order to ensure that the light trajectory lies in the plane $\vartheta=\pi/2$, this being the situation when our expressions for the Schwarzschild region hold true. 

Finally, we get at the observation point $A(t_0)$ for the cross section and $u_t(t_0)/\sqrt{g_{00}(t_0)}$ for the frequency. The redshift is given then by
 \eq{z=\frac{u_t(t_1)\sqrt{g_{00}(t_0)}}{u_t(t_0)\sqrt{g_{00}(t_1)}}-1=\frac{u^t(t_1)\sqrt{g_{00}(t_1)}}{u^t(t_0)\sqrt{g_{00}(t_0)}}-1}{redshift.1}
and the luminosity distance by
\eq{d_L=(1+z)\sqrt{\frac{A(t_0)}{\Omega}}}{lumdist.1}

These expressions complete the determination of the Hubble diagram.

A detailed numerical investigation based on the equations presented is under way.  

\section{Acknowledgement}
The present work was supported by the OTKA grant NI 68228.
 

\appendix
\section{Matching the flat FRW metric to the Schwarzschild metric}\label{Match F-S}

At the bordering spherical surface the line element is
\eq{ds^2=\left(1-\frac{r_g}{r_s}\right)\left[\left(\frac{\partial t_s}{\partial t_f}\right)^2dt_f^2+2\frac{\partial t_s}{\partial t_f}\frac{\partial t_s}{\partial r_f}dt_fdr_f+\left(\frac{\partial t_s}{\partial r_f}\right)^2dr_f^2\right]\nonumber\\
-\;\frac{1}{1-\frac{r_g}{r_s}}\left[\left(\frac{\partial r_s}{\partial t_f}\right)^2dt_f^2+2\frac{\partial r_s}{\partial t_f}\frac{\partial r_s}{\partial r_f}dt_fdr_f+\left(\frac{\partial r_s}{\partial r_f}\right)^2dr_f^2\right]\nonumber\\
-r_s^2\left(d\vartheta^2-\sin^2\vartheta d\varphi^2\right)}{app.A.01}
and the metric is given by
\begin{eqnarray}
g_{00}&=&\left.\left[\left(1-\frac{r_g}{r_s}\right)\left(\frac{\partial t_s}{\partial t_f}\right)^2-\frac{1}{1-\frac{r_g}{r_s}}\left(\frac{\partial r_s}{\partial t_f}\right)^2\right]\right|_{r_f=R_1}=1\label{app.A.02}\\
g_{01}&=&g_{10}=\left.\left[\left(1-\frac{r_g}{r_s}\right)\frac{\partial t_s}{\partial t_f}\frac{\partial t_s}{\partial r_f}-\frac{1}{1-\frac{r_g}{r_s}}\frac{\partial r_s}{\partial t_f}\frac{\partial r_s}{\partial r_f}\right]\right|_{r_f=R_1}=0\quad\label{app.A.03}\\
g_{11}&=&\left.\left[\left(1-\frac{r_g}{r_s}\right)\left(\frac{\partial t_s}{\partial r_f}\right)^2-\frac{1}{1-\frac{r_g}{r_s}}\left(\frac{\partial r_s}{\partial r_f}\right)^2\right]\right|_{r_f=R_1}=-t_f^{\frac{4}{3}}\label{app.A.04}\\
g_{22}&=&\left.-r_s^2\right|_{r_f=R_1}=-t_f^{\frac{4}{3}}R_1^2\label{app.A.05}\\
g_{33}&=&\left.-r_s^2\sin^2\vartheta\right|_{r_f=R_1}=-t_f^{\frac{4}{3}}R_1^2\sin^2\vartheta\label{app.A.06}\\
g_{02}&=&g_{20}=g_{03}=g_{30}=0\label{app.A.07}
\end{eqnarray}
Eqs.(\ref{app.A.05}) and (\ref{app.A.06}) imply
\eq{\left.r_s\right|_{r_f=R_1}=t_f^{\frac{2}{3}}R_1\quad.}{app.A.07.1}
Near the border we can write
\eq{r_s=t_f^{\frac{2}{3}}R_1+\sum_{n=1}^{\infty}(r_f-R_1)^nf_n(t_f)\quad,}{app.A.08}
where the  functions $f_n(t_f)$ for $n>2$ contribute to neither the metric, nor it's derivative at $r_f=R_1$. Thus, we may freely set them to zero, $f_3(t_f)=0,\;f_4(t_f)=0,\;\dots$\;. Continuity of the metric at the border determine  function $f_1(t_f)$, while continuity of the derivatives of the metric at the border determine function $f_2(t_f)$. A similar procedure applies to $t_s$. We expand it in a series near the bordering surface
\eq{t_s=g_0(t_f)+\sum_{n=1}^{\infty}(r_f-R_1)^ng_n(t_f)\quad.}{app.A.09}
The partial derivatives are given by
\eqq{\left.\frac{\partial r_s}{\partial t_f}\right|_{r_f=R_1}&=&\frac{2}{3}t_f^{-\frac{1}{3}}R_1}{app.A.10}
{\left.\frac{\partial r_s}{\partial r_f}\right|_{r_f=R_1}&=&f_1(t_f)}{app.A.11}
{\left.\frac{\partial t_s}{\partial t_f}\right|_{r_f=R_1}&=&\dot{g}_0(t_f)}{app.A.12}
{\left.\frac{\partial t_s}{\partial r_f}\right|_{r_f=R_1}&=&g_1(t_f)}{app.A.13}
Inserting these into Eqs.(\ref{app.A.02}), (\ref{app.A.03}) and (\ref{app.A.04}) we get
\begin{eqnarray}
g_{00}&=&\left.\left[\left(1-\frac{r_g}{r_s}\right)\dot{g}_0^2(t_f)-\frac{1}{1-\frac{r_g}{r_s}}\frac{4}{9}t_f^{-\frac{2}{3}}R_1^2\right]\right|_{r_f=R_1}=1\label{app.A.14}\\
g_{01}&=&g_{10}=\Bigg[\left(1-\frac{r_g}{r_s}\right)\dot{g}_0(t_f)g_1(t_f)\Bigg.\nonumber\\
&&\qquad\qquad\quad\left.\left.-\frac{1}{1-\frac{r_g}{r_s}}\frac{2}{3}t_f^{-\frac{1}{3}}R_1f_1(t_f)\right]\right|_{r_f=R_1}=0\label{app.A.15}\\
g_{11}&=&\left.\left[\left(1-\frac{r_g}{r_s}\right)g_1^2(t_f)-\frac{1}{1-\frac{r_g}{r_s}}f_1^2(t_f)\right]\right|_{r_f=R_1}=-t_f^{\frac{4}{3}}\label{app.A.16}
\end{eqnarray}
Eqs.(\ref{eq:app.A.07.1}), (\ref{app.A.14}) and (\ref{app.A.16}) yield
\eq{\dot{g}_0(t_f)=\frac{t_f^{\frac{1}{3}}}{t_f^{\frac{2}{3}}R_1-r_g}\sqrt{R_1(t_f^{\frac{2}{3}}R_1-r_g+\frac{4}{9}R_1^3)}\quad,}{app.A.17}
and
\eq{f_1(t_f)=\frac{1}{t_f^{\frac{2}{3}}R_1}\sqrt{t_f^2R_1(t_f^{\frac{2}{3}}R_1-r_g)+g_1^2(t_f)(t_f^{\frac{2}{3}}R_1-r_g)^2}\quad,\quad}{app.A.18}
which implies according to Eq.(\ref{app.A.15})
\eq{g_1(t_f)=\frac{2R_1^2t_f}{3(t_f^{\frac{2}{3}}R_1-r_g)}\quad,}{app.A.19}
hence
\eq{f_1(t_f)=\frac{1}{t_f^{\frac{2}{3}}R_1}\sqrt{t_f^2R_1(t_f^{\frac{2}{3}}R_1-r_g+\frac{4}{9}R_1^3)}\quad.}{app.A.20}

Matching conditions follow from continuity of the derivatives of the metric, namely
\eq{\left.\frac{\partial g_{00}}{\partial r_f}\right|_{r_f=R_1}=0}{app.A.21}
\eq{\left.\frac{\partial g_{01}}{\partial r_f}\right|_{r_f=R_1}=\left.\frac{\partial g_{10}}{\partial r_f}\right|_{r_f=R_1}=0}{app.A.22}
\eq{\left.\frac{\partial g_{11}}{\partial r_f}\right|_{r_f=R_1}=0}{app.A.23}
Eq.(\ref{eq:app.A.21}) can be written in the form
\begin{eqnarray}
\left.\frac{\partial g_{00}}{\partial r_f}\right|_{r_f=R_1}&=&\left[\frac{r_g}{r_s^2}\frac{\partial r_s}{\partial r_f}\left(\frac{\partial t_s}{\partial t_f}\right)^2+\left(1-\frac{r_g}{r_s}\right)2\frac{\partial t_s}{\partial t_f}\frac{\partial^2 t_s}{\partial t_f \partial r_f}\right.\nonumber\\
&&\left.\bigg.+\frac{r_g}{(r_s-r_g)^2}\frac{\partial r_s}{\partial r_f}\left(\frac{\partial r_s}{\partial t_f}\right)^2-\frac{1}{1-\frac{r_g}{r_s}}2\frac{\partial r_s}{\partial t_f}\frac{\partial^2 r_s}{\partial t_f \partial r_f}\bigg]\right|_{r_f=R_1}\nonumber\\
&=&\left[\frac{r_g}{r_s^2}\;f_1\;\dot{g}^2_0+\left(1-\frac{r_g}{r_s}\right)2\;\dot{g}_0\;\dot{g}_1+\frac{r_g}{(r_s-r_g)^2}\;f_1\;\frac{4}{9}t_f^{-\frac{2}{3}}R_1^2\right.\nonumber\\
&&\left.\left.-\frac{1}{1-\frac{r_g}{r_s}}\;2\;\frac{2}{3}t_f^{-\frac{1}{3}}R_1\;\dot{f}_1\right]\right|_{r_f=R_1}=0\quad,\label{eq:app.A.24}
\end{eqnarray}
where 
\eqd{\dot{f}_1(t_f)&=&\frac{2t_f^{\frac{2}{3}}R_1-r_g+\frac{4}{9}R_1^3}{3t_f^{\frac{2}{3}}\sqrt{R_1(t_f^{\frac{2}{3}}R_1-r_g+\frac{4}{9}R_1^3)}}}{app.A.25}
{\dot{g}_1(t_f)&=&\frac{6R_1^2(t_f^{\frac{2}{3}}R_1-r_g)-4R_1^3t_f^{\frac{2}{3}}}{9(t_f^{\frac{2}{3}}R_1-r_g)^2}}{app.A.26}
according to Eqs.(\ref{eq:app.A.19}), (\ref{eq:app.A.20}). Hence the Schwarzschild radius is given by
\eq{r_g=\frac{4}{9}R_1^3}{app.A.27}
from Eq.(\ref{eq:app.A.24}). The partial derivatives are given by
\eqt{\left.\frac{\partial r_s}{\partial r_f}\right|_{r_f=R_1}&=&f_1(t_f)=t_f^{\frac{2}{3}}}{app.A.28}
{\left.\frac{\partial t_s}{\partial t_f}\right|_{r_f=R_1}&=&\dot{g}_0(t_f)=\frac{9t_f^\frac{2}{3}}{9t_f^\frac{2}{3}-4R_1^2}}{app.A.29}
{\left.\frac{\partial t_s}{\partial r_f}\right|_{r_f=R_1}&=&g_1(t_f)=\frac{6R_1t_f}{9t_f^{\frac{2}{3}}-4R_1^2}}{app.A.30}
Integrating Eq.(\ref{eq:app.A.29}) we get
\eq{\left.t_s\right|_{r_f=R_1}=t_f+3\frac{r_g}{R_1}t_f^{\frac{1}{3}}-3\left(\frac{r_g}{R_1}\right)^{\frac{3}{2}}\arth\left(t_f^{\frac{1}{3}}\sqrt{\frac{R_1}{r_g}}\right)\quad,}{app.A.31}
therefore
\eq{t_s=t_f+3\frac{r_g}{R_1}t_f^{\frac{1}{3}}-3\left(\frac{r_g}{R_1}\right)^{\frac{3}{2}}\arth\left(t_f^{\frac{1}{3}}\sqrt{\frac{R_1}{r_g}}\right)+\sum_{n=1}^{\infty}(r_f-R_1)^ng_n(t_f)\qquad}{app.A.32}
where functions $g_n(t_f)$ for $n>2$ -- like in case of the coordinate $r_s$  -- do not contribute at $r_f=R_1$. Function $g_1(t_f)$ is already known. Functions  $f_2(t_f)$ and $g_2(t_f)$ can be derived from Eqs.(\ref{eq:app.A.22}), (\ref{eq:app.A.23}), because 
\begin{eqnarray}
\left.\frac{\partial g_{01}}{\partial r_f}\right|_{r_f=R_1}&=&\left.\frac{\partial g_{10}}{\partial r_f}\right|_{r_f=R_1}=\Bigg[\frac{r_g}{r_s^2}\;f_1\;\dot{g}_0\;g_1+\left(1-\frac{r_g}{r_s}\right)(\dot{g}_1\;g_1+\dot{g}_0\;2\;g_2)\Bigg.\nonumber\\
&&\left.\left.+\frac{r_g}{(r_s-r_g)^2}\;f_1\;\dot{f}_0\;f_1-\frac{1}{1-\frac{r_g}{r_s}}(\dot{f}_1\;f_1+\dot{f}_0\;2\;f_2)\right]\right|_{r_f=R_1}\nonumber\\
&=&0\quad,\label{eq:app.A.33}\\\nonumber\\
\left.\frac{\partial g_{11}}{\partial r_f}\right|_{r_f=R_1}&=&\Bigg[\frac{r_g}{r_s^2}\;f_1\;g_1^2+\left(1-\frac{r_g}{r_s}\right)\;2\;g_1\;2\;g_2\Bigg.\nonumber\\
&&\left.\left.+\frac{r_g}{(r_s-r_g)^2}\;f_1\;f_1^2-\frac{1}{1-\frac{r_g}{r_s}}\;2\;f_1\;2\;f_2\right]\right|_{r_f=R_1}\nonumber\\
&=&0\quad,\label{eq:app.A.34}
\end{eqnarray}
hence
\eq{f_2(t_f)=\frac{R_1}{3}\quad,}{app.A.35}
and
\eq{g_2(t_f)=\frac{3t_f\left(9t_f^{\frac{2}{3}}-8R_1^2\right)}{\left(9t_f^{\frac{2}{3}}-4R_1^2\right)^2}\quad.}{app.A.36}
Finally we get the coordinate transformation between the flat FRW and the Schwarzschild metric as follows
\eq{t_s&=&t_f+3\frac{r_g}{R_1}t_f^{\frac{1}{3}}-3\left(\frac{r_g}{R_1}\right)^{\frac{3}{2}}\arth\left(t_f^{\frac{1}{3}}\sqrt{\frac{R_1}{r_g}}\right)\nonumber\\
&&+\left(r_f-R_1\right)\frac{6R_1t_f}{9t_f^{\frac{2}{3}}-4R_1^2}+\left(r_f-R_1\right)^2\frac{3t_f\left(9t_f^{\frac{2}{3}}-8R_1^2\right)}{\left(9t_f^{\frac{2}{3}}-4R_1^2\right)^2}\quad,\quad}{app.A.37}
and
\eq{r_s=t_f^{\frac{2}{3}}r_f+(r_f-R_1)^2\frac{R_1}{3}\quad,}{app.A.38}
together with
\eqd{\vartheta_s&=&\vartheta_f}{app.A.39}
{\varphi_s&=&\varphi_f}{app.A.40}
Note that throughout we used units ensuring $\frac{3}{2}H_0=1$. Return to general units is possible by the  substitutions $t_f \to t_f/t_0$ and $R_1 \to a_0R_1$.

\section{Matching the closed FRW metric to the Schwarzschild metric}\label{Match C-S}

The metric at the bordering surface is given by
\begin{eqnarray}
g_{00}&=&\left.\left[\left(1-\frac{r_g}{r_s}\right)\left(\frac{\partial t_s}{\partial t_c}\right)^2-\frac{1}{1-\frac{r_g}{r_s}}\left(\frac{\partial r_s}{\partial t_c}\right)^2\right]\right|_{r_c=R_2}=1\label{eq:app.B.01}\\
g_{01}&=&g_{10}=\left.\left[\left(1-\frac{r_g}{r_s}\right)\frac{\partial t_s}{\partial t_c}\frac{\partial t_s}{\partial r_c}-\frac{1}{1-\frac{r_g}{r_s}}\frac{\partial r_s}{\partial t_c}\frac{\partial r_s}{\partial r_c}\right]\right|_{r_c=R_2}=0\label{eq:app.B.02}\\
g_{11}&=&\left.\left[\left(1-\frac{r_g}{r_s}\right)\left(\frac{\partial t_s}{\partial r_c}\right)^2-\frac{1}{1-\frac{r_g}{r_s}}\left(\frac{\partial r_s}{\partial r_c}\right)^2\right]\right|_{r_c=R_2}=-{\frac{a^2(t_c)}{1-R_2^2}}\qquad\quad\label{eq:app.B.03}\\
g_{22}&=&\left.-r_s^2\right|_{r_c=R_2}=-a^2(t_c)R_2^2\label{eq:app.B.04}\\
g_{33}&=&\left.-r_s^2\sin^2\vartheta\right|_{r_c=R_2}=-a^2(t_c)R_2^2\sin^2\vartheta\label{eq:app.B.05}\\
g_{02}&=&g_{20}=g_{03}=g_{30}=0\label{eq:app.B.06}
\end{eqnarray}
furthermore the derivatives at the surface are
\eqt{\left.\frac{\partial g_{00}}{\partial r_c}\right|_{r_c=R_2}&=&0}{app.B.07}
{\left.\frac{\partial g_{01}}{\partial r_c}\right|_{r_c=R_2}&=&\left.\frac{\partial g_{10}}{\partial r_c}\right|_{r_c=R_2}=0}{app.B.08}
{\left.\frac{\partial g_{11}}{\partial r_c}\right|_{r_c=R_2}&=&-2a(t_c)^2\;\frac{R_2}{\left(1-R_2^2\right)^2}}{app.B.09}
The coordinates can be written in the following form
\begin{eqnarray}
r_s&=&f_0(\chi)+f_1(\chi)(r_c-\sin\alpha)+f_2(\chi)(r_c-\sin\alpha)^2\\
t_s&=&g_0(\chi)+g_1(\chi)(r_c-\sin\alpha)+g_2(\chi)(r_c-\sin\alpha)^2
\end{eqnarray}
where
\begin{eqnarray}
f_0=\beta \sin^2\chi \sin\alpha
\end{eqnarray}
according to Eq.(\ref{eq:app.B.04}). From Eqs. (\ref{eq:fc.0xd}) and (\ref{eq:app.B.01}) we get
\begin{eqnarray}
\frac{d g_0}{d\chi}=2\beta\frac{\sin^2\chi}{1-\frac{r_g}{f_0}}\sqrt{\sin^2\alpha\cot^2\chi+1-\frac{r_g}{f_0}}\quad,\label{eq:app.B.13}
\end{eqnarray}
and from Eqs.(\ref{eq:app.B.02}) and (\ref{eq:app.B.03}) we have
\begin{eqnarray}
f_1&=&\beta\frac{\sin^2\chi}{\cos \alpha}\sqrt{\sin^2\alpha\cot^2\chi+1-\frac{r_g}{f_0}}\\
g_1&=&\beta\frac{\tan\alpha\sin\chi\cos\chi}{1-\frac{r_g}{f_0}}
\end{eqnarray}
Hence we can express the Schwarzschild radius from Eq.(\ref{eq:app.B.07}) as
\eq{r_g=\beta \sin^3\alpha\quad.}{app.B.16}
Integrating Eq.(\ref{eq:app.B.13}) we get
\begin{eqnarray}
g_0&=&\beta\left[\cos\alpha\left(1+2\sin^2\alpha\right)\chi-\cos\alpha\sin\chi\cos\chi\phantom{\frac{\tan\chi}{\tan\alpha}}\right.\nonumber\\
&&\left.-2\sin^3\alpha\;\arth\left(\frac{\tan\chi}{\tan\alpha}\right)\right]\quad,
\end{eqnarray}
and
\begin{eqnarray}
f_1&=&\beta\sin^2\chi\\
g_1&=&\beta\frac{\tan\alpha\sin^3\chi\cos\chi}{\cos^2\alpha-\cos^2\chi}
\end{eqnarray}
From Eqs.(\ref{eq:app.B.08}) and (\ref{eq:app.B.09}) we have
\begin{eqnarray}
f_2&=&\beta\frac{3}{4}\frac{\sin\alpha}{\cos^2\alpha}\\
g_2&=&\beta\frac{1}{2}\frac{\sin^3\chi\cos\chi\left(\cos^4\alpha-\cos^2\chi\right)}{\cos^3\alpha\left(\cos^2\alpha-\cos^2\chi\right)^2}
\end{eqnarray}
Hence the coordinate transformation at the bordering surface between the closed FRW and the Schwarzschild metric is the following
\eq{t_s&=&\beta\bigg(\cos\alpha\big(1+2\sin^2\alpha\big)\chi-\cos\alpha\sin\chi\cos\chi\bigg.\nonumber\\
&&-\left.2\sin^3\alpha\arth\left(\frac{\tan\chi}{\tan\alpha }\right)+\left(r_c-\sin\alpha\right)\frac{\tan\alpha\sin^3\chi\cos\chi}{\cos^2\alpha-\cos^2\chi}\right.\nonumber\\
&&+\left.\frac{1}{2}\left(r_c-\sin\alpha\right)^2\frac{\sin^3\chi\cos\chi\left(\cos^4\alpha-\cos^2\chi\right)}{\cos^3\alpha\left(\cos^2\alpha-\cos^2\chi\right)^2}\right)\quad,}{4.3.28}
\eq{r_s=\beta\left(r_c\sin^2\chi+\frac{3}{4}(r_c-\sin\alpha)^2\frac{\sin\alpha}{\cos^2\alpha}\right)\quad,}{4.3.47}
together with
\eqd{\vartheta_s=\vartheta_c}{4.3.45}
{\varphi_s=\varphi_c}{4.3.46}


\end{document}